\documentclass[twocolumn,twocolappendix,astrosymb]{aastex631}

\newcommand{\code}[1]{{\texttt{#1}}}

\usepackage{xspace}
\newcommand{\popsed}{\textsc{PopSED}\xspace}
\newcommand{\provabgs}{\code{PROVABGS}}
\newcommand{\md}{\mathrm{d}}

\received{Sept 29, 2023}
\revised{Nov 9, 2023}
\accepted{Nov 10, 2023}

\usepackage{amsmath}
\usepackage{CJKutf8}
\usepackage{bm}
\usepackage{appendix}
\usepackage[caption=false]{subfig}

\shorttitle{Population-Level Inference for Galaxy Properties}
\shortauthors{Li et al.}
\graphicspath{{./}{figures/}}

\begin{document}
\begin{CJK*}{UTF8}{gbsn}

\title{\popsed: Population-Level Inference for Galaxy Properties from Broadband Photometry \\ with Neural Density Estimation}

\correspondingauthor{Jiaxuan Li}
\email{jiaxuanl@princeton.edu}

\author[0000-0001-9592-4190]{Jiaxuan Li (李嘉轩)}
\affiliation{Department of Astrophysical Sciences, 4 Ivy Lane, Princeton University, Princeton, NJ 08544, USA}

\author[0000-0002-8873-5065]{Peter Melchior}
\affiliation{Department of Astrophysical Sciences, 4 Ivy Lane, Princeton University, Princeton, NJ 08544, USA}
\affiliation{Center for Statistics \& Machine Learning, Princeton University, Princeton, NJ 08544, USA}

\author[0000-0003-1197-0902]{ChangHoon Hahn}
\affiliation{Department of Astrophysical Sciences, 4 Ivy Lane, Princeton University, Princeton, NJ 08544, USA}

\author[0000-0003-1385-7591]{Song Huang (黄崧)}
\affiliation{Department of Astronomy and Tsinghua Center for Astrophysics, Tsinghua University, Beijing 100084, China}

\begin{abstract}
We present \popsed, a framework for the population-level inference of galaxy properties from photometric data. Unlike the traditional approach of first analyzing individual galaxies and then combining the results to determine the physical properties of the entire galaxy population, we directly make the population distribution the inference objective. We train normalizing flows to approximate the population distribution by minimizing the Wasserstein distance between the synthetic photometry of the galaxy population and the observed data. We validate our method using mock observations and apply it to galaxies from the GAMA survey. \popsed reliably recovers the redshift and stellar mass distribution of $10^{5}$ galaxies using broadband photometry within $<1$ GPU hr, being $10^{5-6}$ times faster than the traditional spectral energy distribution modeling method. From the population posterior, we also recover the star-forming main sequence for GAMA galaxies at $z<0.1$. With the unprecedented number of galaxies in upcoming surveys, our method offers an efficient tool for studying galaxy evolution and deriving redshift distributions for cosmological analyses.
\end{abstract}

\keywords{Stellar populations (1622) --- Galaxy photometry (611) --- Galaxy evolution (594) --- Neural networks (1933) --- Astrostatistics (1882) --- Sky surveys (1464)}

\section{Introduction}
Galaxies are the building blocks of the Universe. The history of their formation and evolution are encoded in their spectral energy distributions (SEDs). Therefore, one of the most important tasks in extragalactic astronomy is to decode the physical properties of galaxies, including the redshift, stellar mass, star formation history (SFH), and chemical enrichment history, from the observed SEDs. The state-of-the-art SED modeling methods \citep[e.g.,][]{Noll2009,Carnall2018,Johnson2021} that utilize stellar population synthesis (SPS) models \citep{Conroy2013ARAA} have been an indispensable component in many studies, ranging from individual high-redshift galaxies \citep{Labbe2023} to large galaxy surveys \citep[e.g., the Sloan Digital Sky Survey, or SDSS,][]{Gunn2006}. Hundreds of millions of galaxies will be characterized with the upcoming observations from the Rubin Observatory Legacy Survey of Space and Time \citep[LSST;][]{LSST2019}, \textit{Euclid} \citep{Euclid2016}, and the \textit{Roman} telescope \citep{Spergel2015}, enabling a huge discovery space for the origin and evolution of galaxies as a \textit{population}.

SED modeling is typically applied to individual galaxies. The point estimates or the posterior distributions of each galaxy need to be combined to study the galaxy population, such as measuring the stellar mass function \citep[e.g.,][]{Wright2017,Hahn2023SMF}, the star-forming main sequence \citep[SFMS; e.g.,][]{Speagle2014,Thorne2021}, and the stellar mass--metallicity relation \citep[e.g.,][]{Tremonti2004,Curti2020}. However, modeling the SEDs of individual galaxies is very expensive. SED fitting is a high-dimensional problem (typically $>10$ dimensions) that requires evaluating the SPS model and sampling the posterior distribution several million times for a single galaxy. A Bayesian SED fitting with traditional SPS models \citep[e.g., FSPS; ][]{Conroy2009} takes $\sim 20$ CPU hr per galaxy \citep{Leja2019b}, making it computationally formidable to analyze millions of galaxies, not to mention analyze the entire galaxy population. 

This problem has been partially mitigated by recent developments in accelerating SED modeling by emulating the SPS models with neural networks \citep{Alsing2020ApJS}, building differentiable SPS models with a high-performance library \citep{Hearin2021}, and speeding up the sampling by using amortized simulation-based inference \citep[SBI;][]{Hahn2022sedflow,Khullar2022,Wang2023}. However, some of these methods need sophisticated training and are costly to retrain when adapting to different SPS models or noise properties. Furthermore, to constrain the galaxy population distribution, one cannot simply combine point estimates, e.g. the Maximum A Posteriori (MAP) estimate, or directly average the individual posteriors. One needs thousands of samples from the posterior of each galaxy to construct the population posterior by running Markov Chain Monte Carlo (MCMC) for a hierarchical model \citep[e.g.,][]{Malz2020,Alsing2022,Hahn2023SMF}, which poses additional modeling and computational challenges.

In this paper, we introduce \textsc{PopSED}, an efficient and robust method to infer the properties of the galaxy population from broadband photometric data. We directly constrain the population-level distribution of galaxy properties without fitting for individual galaxies and circumvent the need for combining individual posteriors. We use normalizing flows to flexibly model the population distribution, and we train the flows by minimizing the Wasserstein distance between the observed photometric data and the synthetic data generated by the flow. \popsed could reliably recover the population-level properties for $10^5$ galaxies with $\sim 1$ GPU hr. The code is available on GitHub\footnote{\texttt{popsed} codebase: \url{https://github.com/AstroJacobLi/popsed}. under an MIT License and version 0.0.6 is archived in Zenodo \citep{li_2023_10094993}.}

The paper is organized as follows. We describe our method in \S\ref{sec:method} and Figure \ref{fig:flow}, introduce the data we used in \S\ref{sec:data}, validate our method using mock observations and real data in \S\ref{sec:results}, and discuss the strengths and limitations of this work in \S\ref{sec:discuss}. We adopt a \citet{Chabrier2003} initial mass function (IMF) and a flat $\Lambda$CDM cosmology from \citet{Planck15}, with $\Omega_{\rm m}= 0.307$ and $H_0 = 67.7\ $km s$^{-1}$ Mpc$^{-1}$. The photometry used in this work is in the AB system \citep{Oke1983}.

\section{Method}\label{sec:method}
The goal of this paper is to infer the distribution of the physical properties $\bm{\theta}$ of a large number of galaxies from photometric data $\{\bm{X}_i\}$. We approximate the underlying galaxy population distribution $p(\bm{\theta}|\{\bm{X}_i\})$ with a flexible neural density estimator, a normalizing flow $q_{\phi}(\bm{\theta})$ with parameters $\phi$. In order to train the normalizing flow to approximate the population distribution, we compare the synthetic photometric data generated by the flow with the observed data. Specifically, after sampling from the flow $\bm{\theta}_j^{\phi} \sim q_{\phi}(\bm{\theta})$, we predict the corresponding broadband photometry $\hat{\bm{X}}_j^\phi = F(\bm{\theta}_j^\phi)$ using the SPS model. We then compare the observed photometry $\{\bm{X}_i\}$ and the synthetic photometry $\{\hat{\bm{X}}_j^\phi\}$ by calculating the Wasserstein distance $\mathcal{W}_2(\{\hat{\bm{X}}_j^\phi\}, \{\bm{X}_i\})$ between the two distributions. The normalizing flow is trained to minimize $\mathcal{W}_2$ until the synthetic photometry from the normalizing flow agrees with the observed photometry, at which point $q_{\phi}(\bm{\theta})$ serves as a MAP estimate to the galaxy population distribution $p(\bm{\theta}|\{\bm{X}_i\})$. We train an ensemble of flows $\{q_{\phi}(\bm{\theta})\}$ to further approximate the posterior of the population distribution. 

\begin{figure*}
    \centering
    \vbox{
		\centering
		\includegraphics[width=1\linewidth]{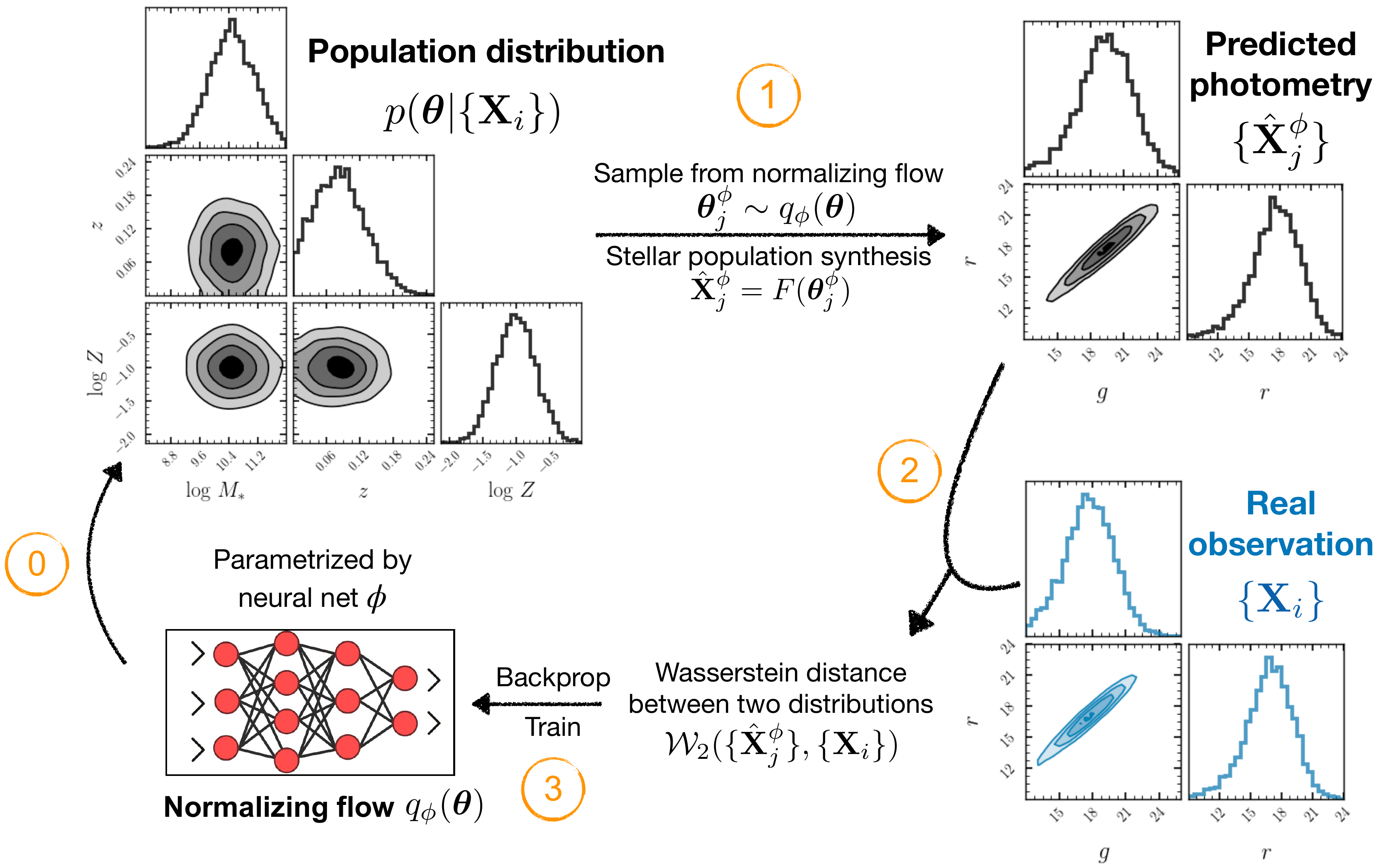}
	}
    \caption{A schematic diagram of \popsed (details in \S\ref{sec:method}). The galaxy population distribution $p(\bm{\theta}|\{\bm{X}_i\})$ is approximated by a normalizing flow $q_{\bm{\phi}}(\bm{\theta})$. We sample from the normalizing flow and forward model the synthetic photometry $\{\hat{\bm{X}}^{\bm{\phi}}_j\}$ using the galaxy SED emulator $F(\bm{\theta}_j^{\phi})$. Then we compare the distributions of the observed photometry and the synthetic photometry by calculating the Wasserstein distance $\mathcal{W}_2(\{\hat{\bm{X}}_j^\phi\}, \{\bm{X}_i\})$, which is used as a loss to train the normalizing flow until the synthetic photometry from the normalizing flow agrees with the observed photometry.
    }
    \label{fig:flow}
\end{figure*}

Figure \ref{fig:flow} provides a high-level overview of our method. We describe the forward model $F$ used to model the photometry in \S\ref{sec:spsmodel} and \S\ref{sec:emulator}, introduce the normalizing flow in \S\ref{sec:nde} and Wasserstein distance in \S\ref{sec:wasserstein}, and present the training strategy in \S\ref{sec:train}.

\subsection{SPS Modeling}\label{sec:spsmodel}

An SPS model is needed to translate the physical parameters of a galaxy $\bm{\theta}_i$ to its SED $\bm{X}_i$ (see \citealt{Conroy2013ARAA} for a review). Different SPS models have different choices for modeling the SFH, chemical enrichment history, and dust attenuation. In the following, we discuss each of these aspects in the SPS model we use.

In this work, the SFHs of galaxies are described using the \provabgs{} model \citep{Alsing2020ApJS,Hahn2022PROVABGS}, which is trained on the galaxies in the Illustris hydrodynamical simulation \citep{Genel2014,Vogelsberger2014,Nelson2015}. The SFH is modeled by a linear combination of four SFH bases $\{s_i^{\rm SFH}\}$ and one burst component $\delta(t-t_{\rm burst})$: 
\begin{equation}\label{eq:sfh_nmf}
    \mathrm{SFH}(t) \propto (1 - f_{\rm burst}) \sum_{i=1}^{4}\beta_i s_i^{\rm SFH}(t) \\
    + f_{\rm burst}\, \delta(t-t_{\rm burst}),
\end{equation}
where $\beta_i$ is the coefficient of each SFH base, $t_{\rm burst}$ is the lookback time when the starburst happens, and $f_{\rm burst}$ is the fraction of the total stellar mass that is formed during the burst. The SFH bases $\{s_i^{\rm SFH}\}$ are generated based on the SFHs of Illustris galaxies using the non-negative matrix factorization \citep[][]{Lee1999,Cichocki2009}. As shown in \citet{Hahn2022PROVABGS}, the four bases are sufficient to capture the SFH of Illustris galaxies. Among the SFH bases, $s_1^{\rm SFH}$ corresponds to the most recent star formation, whereas $s_4^{\rm SFH}$ corresponds to the oldest star formation (see Fig. 5 and Appendix A in \citealt{Hahn2022PROVABGS}). Such SFH bases are non-negative by construction and are physically intuitive to interpret. We require $\sum_i \beta_i = 1$, therefore only three of the $\{\beta_i\}$ are independent\footnote{In practice, we sample $\beta_i$ by sampling three independent uniform distributions; see Appendix \ref{ap:sps} for details.}. We define the total formed stellar mass to be the integration of SFH: $M_{\star,\rm formed} = \int_{t=0}^{t_{\rm age}} \mathrm{SFH}(t, t_{\rm age})$. Unless otherwise noted, in this work we refer to the stellar mass $M_\star$ as the total formed stellar mass $M_{\star,\rm formed}$\footnote{We note that some surveys report the stellar mass as the total mass of all luminous material at the time of observation (also called the ``surviving stellar mass''). The difference in the definition of stellar mass should be noticed when comparing the results. \citet{Hearin2021} provided a fitting function to calculate the surviving stellar mass fraction.}. In total, the SFH of a galaxy is described by 7 parameters, namely $\beta_1, \beta_2, \beta_3, \beta_4$, $f_{\rm burst}$, $t_{\rm burst}$, and $M_\star$.

Another important piece in the SPS model is the chemical enrichment history, which is usually simplified to be the metallicity history (ZH) by assuming a fixed element abundance ratio for all metals. In the original \provabgs{}, the ZH is described as a linear combination of two ZH bases extracted from Illustris. Because photometric data alone are quite uninformative in inferring the ZH \citep[e.g.,][]{Hahn2022PROVABGS}, we simplify the ZH to be a constant stellar metallicity $\log(Z_\star/Z_\odot)$ over time, as many other works assume \citep[e.g.,][]{Carnall2019,Leja2019}. 

In order to synthesize the stellar populations, we discretize the lookback time $t$ into time bins. The first time bin corresponds to $\log\, (t/\mathrm{yr}) < 6.05$, and each time bin has a width of 0.1 dex until the lookback time reaches the age of the galaxy $t_{\rm age}$, which is determined by its redshift $z$. We treat the stellar population in each time bin as a simple stellar population (SSP) and we evaluate its SFR according to the SFH. In the end, we add the spectra of the SSPs in each time bin together weighted by the stellar mass formed in each bin. We assume a \citet{Chabrier2003} IMF and use the MIST isochrones \citep{Choi2016,Dotter2016} and the empirical spectra library MILES \citep{MILES} for 3800--7100\AA{} and the BaSeL library \citep{Lejeune1997,Lejeune1998,Westera2002} for wavelengths outside of the range of MILES. We use FSPS \citep{Conroy2009,Conroy2010,pyfsps} to generate SSP spectra. We do not add nebular emissions to the spectra because broadband photometry reflects more about the continuum of the galaxy spectrum.

We add dust attenuation to the galaxy spectra following the recipe in \citet{Charlot2000}, which includes the birth-cloud attenuation and the diffuse dust screening. The birth-cloud attenuation only acts for stars younger than $10^7$ yr \citep{Conroy2009}, whereas the diffuse dust component affects all stars. We refer interested readers to \citet{Hahn2022PROVABGS} for details. There are three parameters in our dust model: $\tau_1$ is the birth-cloud optical depth at 5500\AA, $\tau_2$ is the diffuse dust optical depth at 5500\AA, and $n_{\rm dust}$ is the slope of the \citet{Calzetti2000} attenuation curve. 

To summarize, our SPS model contains 12 parameters, as listed in Table \ref{tab:parameters}. We refer the interested readers to \citet{Hahn2022PROVABGS} for a more complete description of \provabgs{}. We emphasize that our method for population-level inference is not limited to specific SPS models. One can use a different SPS model to perform the population-level inference on galaxy populations. If the SPS model is relatively slow to evaluate, one needs to train a neural emulator to accelerate the evaluation and bring differentiability to the inference, as we introduce below. 

\begin{deluxetable*}{p{0.1\textwidth} p{0.4\textwidth} p{0.4\textwidth}}
\tablecaption{Parameters in the SPS model and their priors for training the galaxy spectrum emulator \label{tab:parameters}}
\tablehead{
\colhead{Parameter} & \colhead{Description} & \colhead{Prior for Training the Emulator}
}
\startdata
\hline
$z$ & Redshift & $\rm{Uniform}\,(0,\ 1.55)$\\
$\log(M_{\star}/M_{\odot})$ & log10 stellar mass & Fixed to $M_\star = 1\ M_\odot$ when training the emulator\\
$\beta_1, \beta_2, \beta_3, \beta_4$ & Coefficients of SFH bases (Eq.~\ref{eq:sfh_nmf}) & Flat Dirichlet prior with $0 \leqslant \beta_i \leqslant 1,\ \sum_i \beta_i = 1$ (see Appendix \ref{ap:sps})\\
$t_{\rm burst}$ [Gyr] & The lookback time when the star formation burst happens (Eq.~\ref{eq:sfh_nmf}) & $\rm{Uniform}\,(10^{-2},\, 13.27)$ \\
$f_{\rm burst}$ & The fraction of total stellar mass formed in the star formation burst (Eq. \ref{eq:sfh_nmf}) & $\rm{Uniform}\,(0,\, 1)$ \\
$\log(Z_{\star}/Z_{\odot})$ & Stellar metallicity ($Z_\odot=0.019$) & $\rm{Uniform}\,(-2.6,\ 0.3)$ \\
$n_{\rm dust}$ & The power-law index of the \citet{Calzetti2000} attenuation curve & $\rm{Uniform}\,(-3.0,\ 1.0)$\\
$\tau_{\rm 1}$ & Birth-cloud dust optical depth & $\rm{Uniform}\,(0,\ 3.0)$ \\
$\tau_{\rm 2}$ & Diffuse dust optical depth & $\rm{Uniform}\,(0,\ 3.0)$ \\
\hline
\enddata
\end{deluxetable*}

\subsection{Galaxy Spectrum Emulator}\label{sec:emulator}

Although we do not use MCMC to construct the population distribution in this work, we also need a fast and differentiable SPS model, such that the information contained in the photometric data (i.e., the Wasserstein distance; see \S\ref{sec:wasserstein}) could efficiently flow back to the parameter space. Therefore, we train an emulator for the SPS model described in \S\ref{sec:spsmodel} following the approach in \citet{Alsing2020ApJS} and \citet{Hahn2022PROVABGS}. After training, the emulator takes the physical parameters $\bm{\theta}_i$ of a galaxy (listed in Table \ref{tab:parameters}) and predicts its synthetic photometry $\hat{\bm{X}}_i = F(\bm{\theta}_i)$ in SDSS $ugriz$ bands with realistic noise added. We choose SDSS bands just to better match with the data used in \S\ref{sec:data}. One can certainly include more filters if needed. The details of training the spectrum emulator are presented in Appendix \ref{ap:sps}.

Noise must be added to the noiseless SEDs in order to meaningfully forward model the observed data. For simplicity, we assume the noise in each filter is independent and construct the noise model in each band $k$. To be specific, we add Gaussian noise to the noiseless fluxes $f_k$ following $\hat{f}_k = f_k + n_k$, where $n_k \sim \mathcal{N}(0, \sigma_k)$. The noise level $\sigma_k$ should depend on the flux $f_k$ because, given a survey depth, fainter sources have a lower signal-to-noise ratio ($\mathrm{SNR}_k = f_k / \sigma_k$). We assume that the SNR for a given flux $f_k$ follows a Gaussian distribution: $p(\mathrm{SNR}_k|f_k) = \mathcal{N}(\mu_{\rm SNR}(f_k), \sigma_{\rm SNR}(f_k))$. For a given data set, we empirically estimate the median SNR ($\mu_{\rm SNR}$) and the standard deviation ($\sigma_{\rm SNR}$) as a function of $f_k$ in each band $k$ by evaluating them in magnitude bins and interpolating over the bins. During the forward modeling, we sample the distribution $\mathrm{SNR}_k \sim p(\mathrm{SNR}_k|f_k)$ for a given $f_k$, convert $\mathrm{SNR}_k$ to $\sigma_k = f_k / \mathrm{SNR}_k$, then add Gaussian noise $n_k \sim \mathcal{N}(0, \sigma_k)$ to the noiseless flux $f_k$. If the resulting flux $\hat{f}_k < 0$, we repeat the above procedure until $\hat{f}_k \geqslant 0$. Unlike in \citet{Hahn2022sedflow}, who model $p(\sigma_k|f_k)$, we find that modeling $p(\mathrm{SNR}_k|f_k)$ is more robust in the low-SNR regime. In the end, we convert noisy fluxes $\hat{\bm{f}_i}$ to magnitudes $\hat{\bm{X}_i}$. Our emulator is $\sim 10^{3-4}$ times faster than the direct SPS computation.

\subsection{Neural Density Estimator}\label{sec:nde}
We try to approximate the population distribution $p(\bm{\theta}|\{\bm{X}_i\})$ using density estimators. Although traditional density estimation techniques, such as Gaussian mixture models \citep[e.g.,][]{Bovy2011}, are easy to optimize and interpret, they are less flexible and scalable to larger data sets in higher dimensions than the neural density estimation (NDE) techniques. In this work, we use ``normalizing flows'' \citep[e.g.,][]{Tabak2010,Tabak2013,Kobyzev2019} to approximate the population distribution of galaxy properties. Being a neural density estimator, normalizing flows have been successfully applied to many fields in astronomy \citep[e.g.,][]{Alsing2019,Zhang2021,Ciuca2022,Dai2022,Hahn2022sedflow,Green2023} to approximate density distributions and generate new data.

A normalizing flow maps a complex distribution $q_{\phi}(\bm{\theta})$ to a simple base distribution $\pi(\bm{z})$ using an invertible bijective transformation $g: \bm{z} \rightarrow \bm{\theta}$, which is described by a neural network with parameters ${\phi}$. The base distribution is often chosen to be easy to evaluate and sample, such that we can evaluate the target distribution following 
$$q_{{\phi}}(\bm{\theta}) = \pi(g^{-1}(\bm{\theta}))\left|\det\left(\frac{\partial g^{-1}}{\partial \bm{\theta}}\right)\right|.$$
From many different flow models \citep[e.g.,][]{Papamakarios2017}, we use the Neural Spline Flow \citep[NSF; ][]{Durkan2019}, where the base function is a multivariate Gaussian distribution and the transformations are described by monotonic rational quadratic splines. The flexibility of the NSF model makes it well suited to modeling the population distribution. We use the NSF implementation in the \code{sbi}\footnote{\url{https://github.com/mackelab/sbi}} package \citep{Greenberg2019, SBI}. Our NSF has 20 blocks, 60 bins for the splines, and 100 latent features in each coupling layer. 

The original NSF model is initialized such that $q_{\bm{\phi}}(\bm{\theta})$ is a standard multivariate Gaussian distribution. However, such an initialization poses significant challenges for training the flow. First, nonphysical parameters (e.g., negative redshift and stellar mass) or parameters that lie outside the prior of the emulator (Table \ref{tab:parameters}) will be drawn from this initial distribution. These nonphysical values are either meaningless or incompatible with our forward model. Second, the standard Gaussian initialization imposes a quite strong prior, which could bias the population distribution. 

To mitigate these issues, we opt for uniform distributions over the standard Gaussian to initialize the normalizing flow. We append an additional layer to the NSF flow that performs cumulative distribution function (CDF) transformation. The CDF transformation converts a standard Gaussian distribution into a uniform distribution within a user-specified range. We set the ranges of the uniform distributions following the priors listed in Table \ref{tab:parameters}. In particular, for the results shown in \S\ref{sec:results}, we narrow the redshift range to $0 < z < 0.8$ and the stellar mass range to $7.5 < \log(M_\star/M_\odot) < 13.0$ to remove irrelevant parameter space and make the inference more efficient. After applying this CDF transformation, the initial normalizing flow describes a uniform distribution in all dimensions and ensures that all parameters drawn from the initial normalizing flow are physical. The uniform distribution serves as the prior of our inference for the population distribution. 

\subsection{Wasserstein Distance}\label{sec:wasserstein}
The job now is to train the normalizing flow $q_{\phi}(\bm{\theta})$ to best approximate the population distribution. We use optimization to find a flow that is most probable to produce the photometric data that is consistent with the observations. This is equivalent to finding the MAP estimate to the posterior of the population distribution. 
In order to do so, we sample from the normalizing flow $\bm{\theta}_j^{\phi} \sim q_{\phi}(\bm{\theta})$ and generate corresponding synthetic photometry using the forward model $\hat{\bm{X}}_j^{\phi} = F(\bm{\theta}_j)$. The \textit{distance} (or divergence) between the distribution of the observed data $\{\bm{X}_i\}$ and the distribution of the synthetic data $\{\hat{\bm{X}}_j^{\phi}\}$ can be a proxy for the dissimilarity between the normalizing flow and the underlying population distribution. By minimizing this distance metric, we train the normalizing flow \( q_{\phi}(\bm{\theta}) \) so that it accurately approximates the underlying population distribution.

Choosing an appropriate distance metric for probability distributions is critical. However, traditional distance metrics are challenging to apply to high-dimensional and discrete data sets. Because we try to compare two discrete samples with different sizes ($\{\hat{\bm{X}}_j^{\phi}\}$ and $\{\bm{X}_i\}$), the commonly used Kullback-Leibler (KL) divergence cannot be employed without modeling the two distributions separately.

On the other hand, the divergences based on the optimal transport theory, such as the Wasserstein distance (also known as the earth mover's distance), are shown to be well behaved when the KL divergence is not applicable. The Wasserstein distance quantifies the minimal ``total cost'' required to move a distribution (which can be viewed as a volume of soil) to the target distribution (the other volume of soil). This distance metric allows one to compare discrete distributions whose supports do not overlap and to quantify the spatial shift between them. Importantly, the Wasserstein distance is differentiable by construction, symmetric under exchange of the two distributions, and satisfies the triangle inequality, making it well suited for comparing two discrete photometric distributions. Wasserstein distance has been successfully applied to various topics in statistics, including variational inference \citep{Ambrogioni2018}, approximate Bayesian computation \citep{Bernton2019}, and, most famously, Generative Adversarial Networks \citep{Arjovsky2017}. It is also used in astronomy to characterize galaxy images \citep{Holzschuh2022} and the topology of large-scale structures \citep{Tsizh2023}. We refer interested readers to \citet{Peyre2018} and \citet{geomloss} for more details on the Wasserstein distance and its application. 

Practically, the optimal transport solution is approximated using the Sinkhorn algorithm \citep{Sinkhorn,Cuturi2013}, which provides an efficient and scalable approximation to the Wasserstein distance. The Sinkhorn algorithm tries to minimize the earth moving cost together with an entropic regularization term with a coefficient $\varepsilon$ (also denoted as ``temperature''). Such a regularization makes the optimal transport problem solvable using iterations that only involve linear algebra and can run in parallel on GPUs with differentiablility. Physically speaking, the particles in the base distribution are mapped to a fuzzy collection of the target particles whose diameters are proportional to a blurring scale $\sigma = \varepsilon^{1/p}$ \citep[][where $p$ is the index of the $L^{p}$ norm used to calculate the cost]{geomloss}. Thus, the Sinkhorn distance converges to the Wasserstein distance as $\sigma \to 0$. A larger $\sigma$ will produce a fuzzier match between the two distributions but a faster convergence. 

In this work, we use the Wasserstein-2 distance ($\mathcal{W}_2$, where the $L^2$ norm is used in the cost function) to characterize the distance between the observed photometry $\{\bm{X}_i\}$ and the synthetic photometry $\{\hat{\bm{X}_j^{\phi}}\}$. $\mathcal{W}_2$ is calculated using the implementation of the Sinkhorn iteration in the Python package \code{GeomLoss}\footnote{\url{https://www.kernel-operations.io/geomloss/api/install.html}} \citep{Ramdas2017,geomloss,pykeops}. In our case, the two photometric data sets are often not of the same size. \code{GeomLoss} could nicely handle this unbalanced optimal transport problem. The Wasserstein distance $\mathcal{W}_2(\{\hat{\bm{X}}_j^{\phi}\}, \{\bm{X}_i\})$ is then used as a loss to train the normalizing flow $q_{\phi}(\bm{\theta})$.

\subsection{Training}\label{sec:train}
During each training iteration, synthetic photometry is generated by sampling from the normalizing flow, and the Wasserstein distance $\mathcal{W}_2(\{\hat{\bm{X}}_j^{\phi}\}, \{\bm{X}_i\})$ is computed. The gradient of this distance metric is then back-propagated to train the normalizing flow. Because we initialize the normalizing flow to be a uniform distribution, the distribution of the synthetic photometry is quite far from the observed one in the first few steps. Therefore, we calculate the Wasserstein distance using a relatively large blurring scale $\sigma$ at the beginning of the training to capture the global structure of the population distribution. Then we take an annealing strategy to gradually reduce $\sigma$ such that the focus of the normalizing flow shifts from matching the mean values to matching increasingly subtle details as the training proceeds \citep{Chui2000}. Compared with using a small $\sigma$ for all iterations, training using annealing $\sigma$ is faster to converge. In practice, we initialize $\sigma$ to be 0.3 ($\sigma$ corresponds to the \code{blur} parameter in \code{GeomLoss}) and decrease $\sigma$ by 0.05 every 60 steps until reaching $\sigma=0.05$, after which we fix $\sigma=0.002$ for the remaining iterations. We find that the specific annealing schedule does not significantly affect the outcomes.

The realistic noise added to the synthetic photometry $\hat{\bm{X}}^{\bm{\phi}}_j$ can make the training much harder at the beginning, when the normalizing flow has not yet learned the global landscape of the population distribution. Therefore, we take an ``anti-annealing'' strategy, where the noise (described by an effective SNR) is added gradually as the training goes on. In this way, the normalizing flow will be guided by the bulk of the data in the beginning, without paying much attention to the noise. The flow is then exposed to more realistic noises later on and adjusts its shape to match the details in the observed data. To do so, we reduce the noise level $\sigma_k$ in the forward model (see \S\ref{sec:emulator}) by a factor of $R$, which follows an exponential decay with time: $R(t) = 1 + R_0 \cdot \exp(-\tau \cdot t/T)$, where $\tau$ is the decay rate and $T$ is the total epochs of training. For the case studies in \S\ref{sec:results}, we choose to use $R_0=30,\ \tau=12$, and $T=800$. We determine these values by trial and error based on the mock test (see \S\ref{sec:mock}).

\begin{deluxetable*}{ll}
\tablecaption{The distribution of SPS parameters for the mock galaxy population \label{tab:mock}. $\kappa_j$ is used to generate $\beta_i$ of the SFH; see \S\ref{sec:spsmodel} and Appendix \ref{ap:sps}.}
\tablehead{
\colhead{Parameter} & \colhead{Distribution}
}
\startdata
\hline
$z$ and $\log(M_{\star}/M_{\odot})$ & Follow the joint distribution from GAMA DR3 data\\
$\kappa_1$ & Truncated normal: $\mathrm{min}=0$, $\mathrm{max}=1$, $\mu=0.5$, $\sigma=0.3$\\
$\kappa_2, \kappa_3$ & $\rm{Uniform}\,(0,\ 1)$\\
$t_{\rm burst}$ [Gyr] & Truncated normal: $\mathrm{min}=10^{-2}$, $\mathrm{max}=13.27$, $\mu=12$, $\sigma=7$\\
$f_{\rm burst}$ & Truncated normal: $\mathrm{min}=0$, $\mathrm{max}=1$, $\mu=0.1$, $\sigma=0.7$\\
$\log(Z_{\star}/Z_{\odot})$ & Truncated normal: $\mathrm{min}=-2.6$, $\mathrm{max}=0.3$, $\mu=-1.2$, $\mu=0.9$\\
$n_{\rm dust}$ & Truncated normal: $\mathrm{min}=-3.0$, $\mathrm{max}=1.0$, $\mu=2$, $\sigma=2$ \\
$\tau_{\rm 1}$ & Truncated normal: $\mathrm{min}=0$, $\mathrm{max}=3.0$, $\mu=1$, $\sigma=0.8$ \\
$\tau_{\rm 2}$ & Truncated normal: $\mathrm{min}=0$, $\mathrm{max}=3.0$, $\mu=0.6$, $\sigma=0.8$ \\
\hline
\enddata
\end{deluxetable*}

A trained normalizing flow is one MAP solution to the population distribution. It is also critical to understand the \textit{posterior} of the population distribution, i.e., the variation of the MAP solutions. The deep ensemble method is a popular approach to characterizing the Bayesian posterior by training the same deep-learning model multiple times with different initializations and averaging the resulting models. The deep ensemble method is proved to better approximate the Bayesian posterior than methods such as variation inference \citep[e.g.,][]{Wilson2020}. 
We take this ensemble learning approach and train a number of normalizing flows with different random seeds, then combine these models by drawing the same number of samples from them and aggregating the samples. The ensemble of flows would roughly approximate the posterior of the population distribution. 



\popsed is implemented in \code{PyTorch} \citep{pytorch} and trained with the stochastic gradient descent optimizer \code{Adam} \citep{Adam}. We use a one-cycle learning-rate policy \citep{Smith2017} with an initial learning rate of $3\times 10^{-5}$, a maximum learning rate of $3\times 10^{-4}$, and a minimum learning rate of $3\times 10^{-6}$. The learning rate increases from the initial learning rate to the maximum learning rate, and then slowly drops to a minimum learning rate. We train each normalizing flow for 800 epochs. The training loss and the validation loss are comparable, indicating that the flows do not overfit. 


\begin{figure*}
\fig{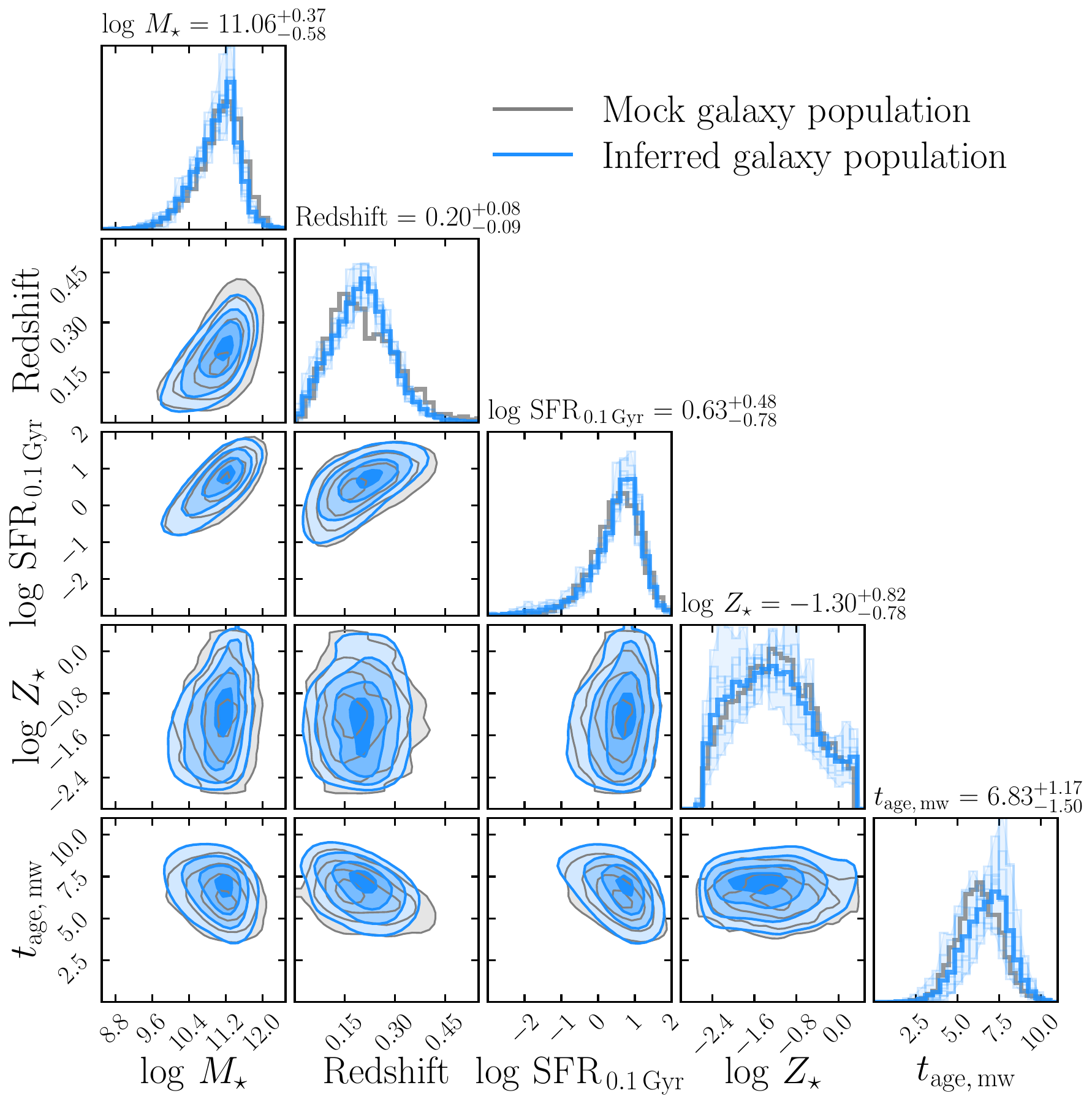}{0.49\linewidth}{}
\fig{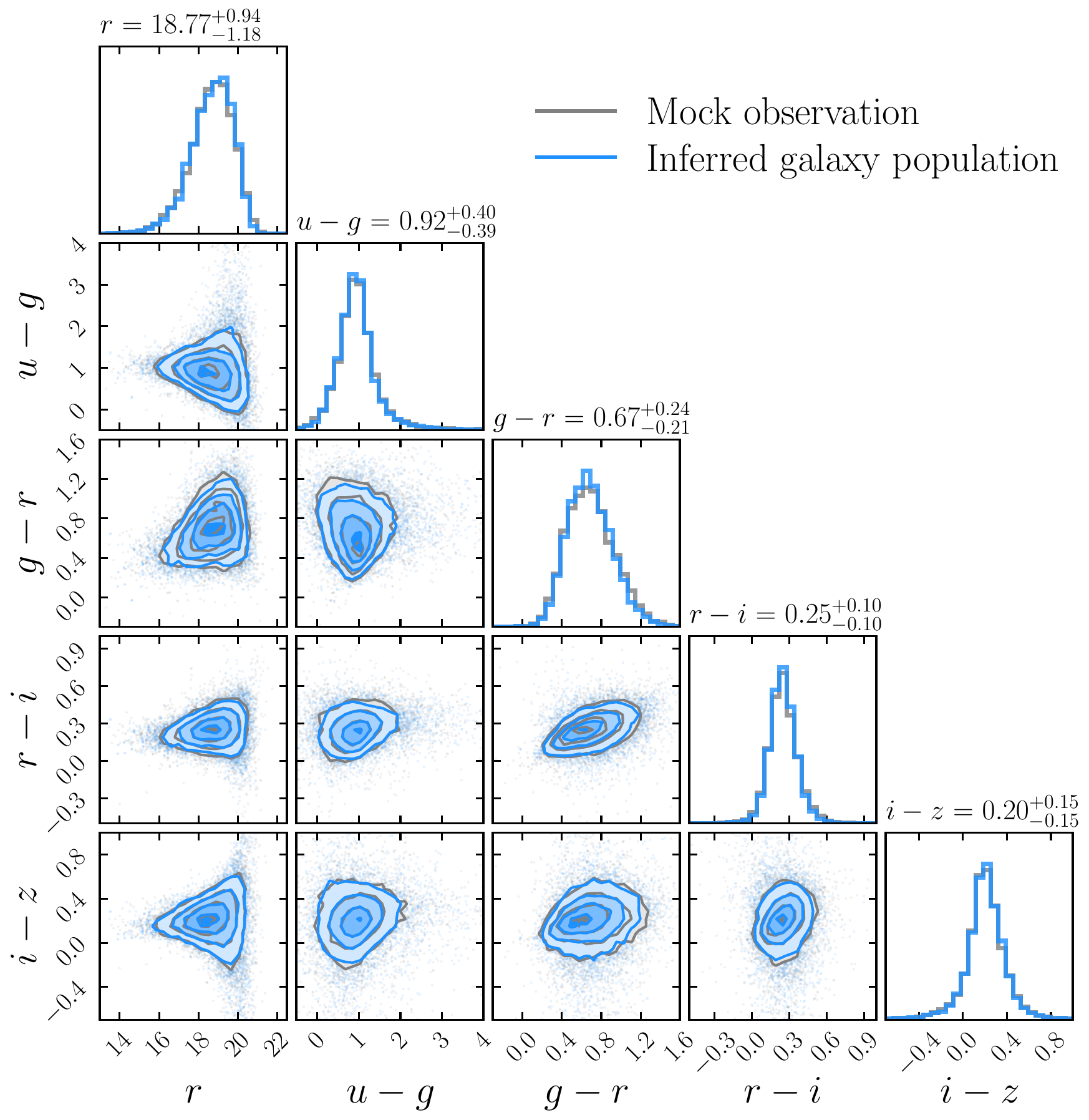}{0.49\linewidth}{}
\vspace{-1em}
 \caption{\textit{Left}: the mock galaxy population (gray contours) and the inferred galaxy population (blue contours) using our method. We calculate the average SFR within the past 0.1 Gyr ($\log\rm{SFR}_{0.1 Gyr}$) and the mass-weighted age ($t_{\rm age, MW}$) using the inferred SPS parameters. The lighter blue histograms show the individual normalizing flows, and the dark blue histogram is the result after averaging 10 flows. The inferred galaxy population agrees with the truth very accurately. \textit{Right}: the mock photometric data in the SDSS $ugriz$ bands (gray contours) are practically indistinguishable from the photometry of the inferred galaxy population using \popsed (blue contours). 
 \label{fig:mock_params_phot} 
 }
\end{figure*}

\section{Data}\label{sec:data}
To test the performance of \popsed on inferring the galaxy properties and redshifts, we use the photometric data used in the Galaxy And Mass Assembly (GAMA) survey \citep{Driver2011}. The GAMA survey is a spectroscopic survey targeting galaxies selected from photometric surveys down to $r<19.8$ mag. It is therefore an ideal data set to test how well we can recover the redshift distribution by comparing our result with the spectroscopic redshifts. We use the aperture-matched photometry \citep{Driver2016} from GAMA Data Release 3 (DR3;\footnote{\url{http://www.gama-survey.org/dr3/}} \citealt{Baldry2018}) in this work. The photometry in the $ugriz$ bands comes from SDSS DR7 \citep{SDSS-DR7}. We remove objects lacking the \code{AUTO} photometry (i.e., Kron photometry) and objects with no flux uncertainty. We also apply a color cut $(J-K_s) > 0.025$ using photometry from the VIKING survey \citep{VIKING2013} for star--galaxy separation and an additional signal-to-noise ratio cut $\mathrm{SNR} > 1$ for all SDSS $ugriz$ bands to remove marginally detected sources with poor photometric quality. There are 83,692 objects in our GAMA sample. Their spectroscopic redshift distribution peaks around $z=0.15$ and extends from $z=0$ to $z\sim 0.60$. In this work, we only use the \code{AUTO} photometry in the SDSS $ugriz$ bands to infer the population distribution. We also build the noise models for each SDSS band as described in \S\ref{sec:emulator}.

\section{Results}\label{sec:results}
With the \popsed framework presented above, we first test how well it works by applying it to mock observations (\S\ref{sec:mock}), where the ground truth is known. Then we apply our method to the GAMA data and compare the inferred properties with spectroscopic results and literature in \S\ref{sec:gama}.

\begin{figure*}
\fig{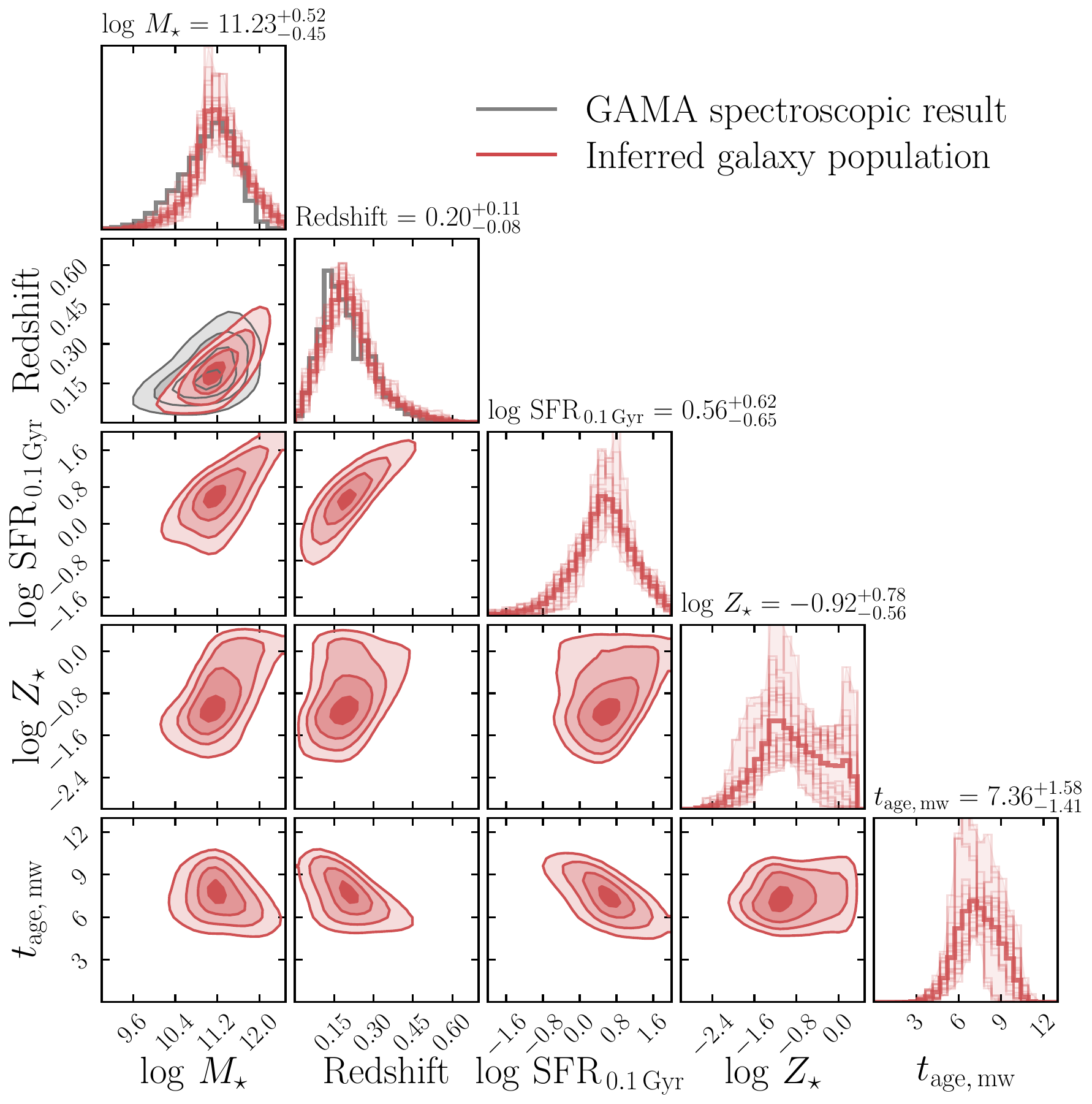}{0.49\linewidth}{}
\fig{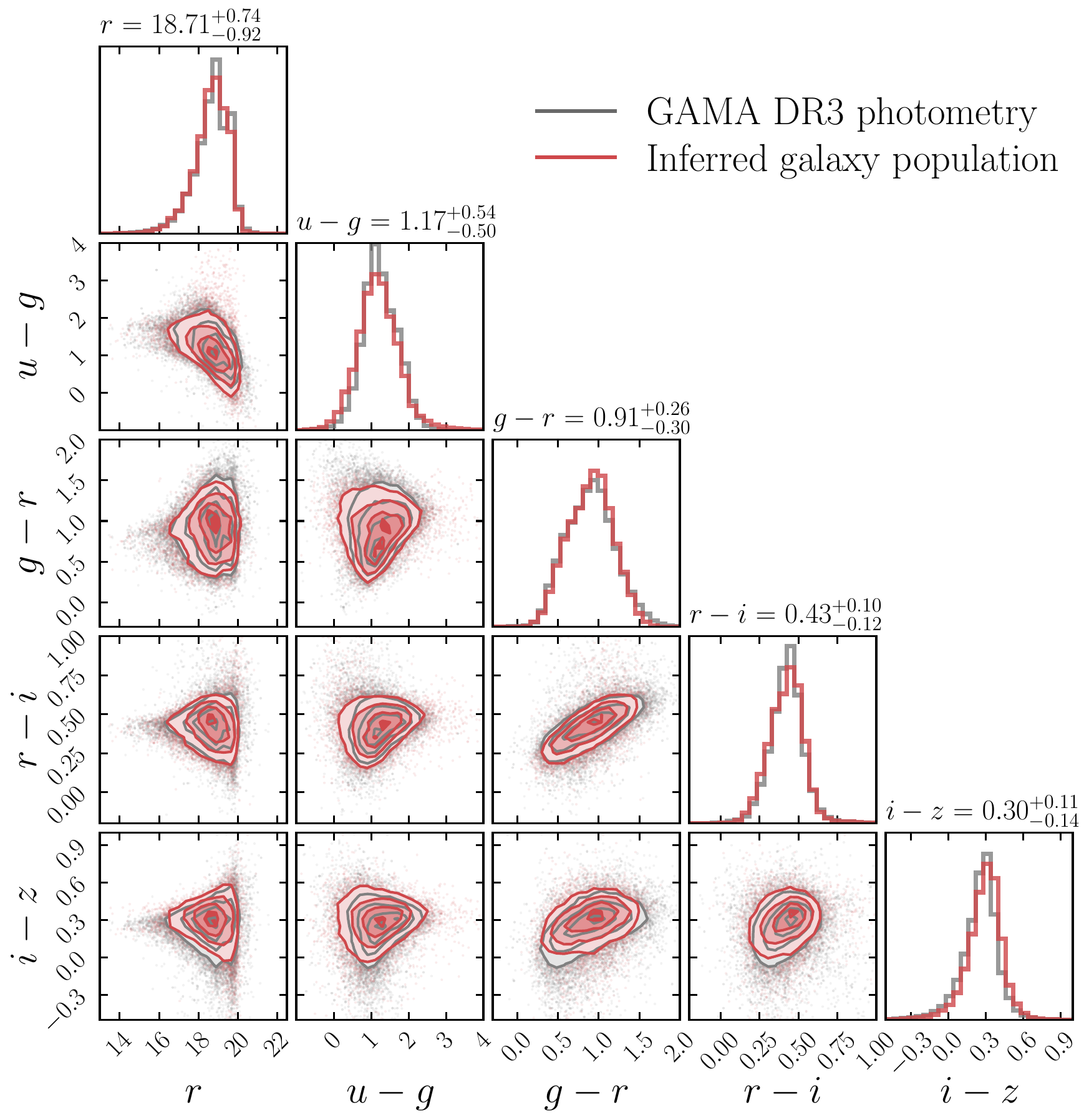}{0.49\linewidth}{}
\vspace{-1em}
 \caption{\textit{Left}: the inferred galaxy population from GAMA photometric data (red contours) and the distribution of spectroscopic redshift and stellar mass from the GAMA catalog (gray contours). The inferred redshift and stellar mass distributions agree well with the spectroscopic results. \textit{Right}: the GAMA photometric data (gray contours) and the synthetic photometry of the inferred galaxy population (red contours). Despite the small difference in the $u-g$ color, the two photometric distributions agree with each other quite well. 
 \label{fig:gama_params_phot} 
 }
\end{figure*}

\subsection{Mock Observations}\label{sec:mock}
We construct a mock observation to test our method and understand its strengths and limitations. We first design a parameter distribution $p(\bm{\theta})_{\rm mock}$ that roughly imitates a real galaxy population in the Universe, then sample from it and generate realistic synthetic photometry as the mock data set. Acknowledging the fact that the redshift and stellar mass should be correlated due to the depth limit of the survey, we use the GAMA DR3 spectroscopic results as a basis and resample the joint distribution of stellar masses and spectroscopic redshifts from GAMA DR3. The distributions of other parameters are listed in Table \ref{tab:mock}. For simplicity, parameters other than $\log(M_\star/M_\odot)$ and $z$ are assumed to be independent. Subsequently, we generate mock observations $\bm{X}_i=F(\bm{\theta}_i)$ in the SDSS $ugriz$ bands for $\bm{\theta}_i \sim p(\bm{\theta})_{\rm mock}$. Realistic noise is added according to the noise model of GAMA DR3 (see \S\ref{sec:data}). In total, our mock galaxy sample comprises 100,863 galaxies with $\rm{SNR}>1$ across all SDSS bands.

To better visualize the mock galaxy population, we calculate several characteristic quantities of galaxies, including the average SFR in the past 0.1 Gyr ($\log\rm{SFR}_{0.1 Gyr}$) and the mass-weighted age ($t_{\rm age, MW}$), from the parameter distribution $p(\bm{\theta})_{\rm mock}$. The gray contours in Figure \ref{fig:mock_params_phot} represent the mock galaxy population by showing the joint distributions of stellar mass, redshift, SFR, metallicity, and age. 
The corresponding distributions of photometry data are also displayed as gray contours in the right panel of Figure \ref{fig:mock_params_phot}. We plot the color--magnitude and color--color distributions for better visibility, although we emphasize that \popsed is trained on magnitudes instead of colors. The distribution of the mock galaxy population in the full parameter space is shown in Figure \ref{fig:mock_posterior} in Appendix \ref{ap:posterior}.

We run \popsed on the mock photometric data of 100,863 galaxies. Each normalizing flow takes $\sim 40$ minutes for training using an NVIDIA A100 GPU. We train 10 independent normalizing flows with different random seeds and aggregate the samples drawn from each flow. The inferred posteriors of the population distribution are shown as the blue contours in the left panel of Figures \ref{fig:mock_params_phot} and \ref{fig:mock_posterior} in Appendix \ref{ap:posterior}. The light blue histograms indicate the variations among different flows, whereas the dark blue histograms show the combined results. We find that photometric data could constrain the stellar mass, redshift, and SFR of the galaxy population very accurately. The inferred galaxy population agrees with the ground truth (gray contours) remarkably well in all dimensions. Moreover, the ensemble of flows encompasses the true distribution, suggesting that \popsed is able to approximate the posterior of the population distribution using the ensemble method. However, the stellar metallicity and mass-weighted age have relatively large scatters among different flows, but we understand this as the photometric data not being informative enough to constrain the distributions of these two parameters.

\subsection{GAMA Sample} \label{sec:gama}
Motivated by the success of the mock test in \S\ref{sec:mock}, we apply \popsed to 83,692 galaxies in the GAMA sample (see \S\ref{sec:data}). We run 30 independent normalizing flows and combine their samples together. The left panel in Figure \ref{fig:gama_params_phot} shows the inferred population distribution among the stellar mass, redshift, average SFR within the past 0.1 Gyr, stellar metallicity, and the mass-weighted age. Similar to Figure \ref{fig:mock_params_phot}, the light red histograms correspond to individual flows.
For completeness, we also show the full population distribution in Figure \ref{fig:gama_posterior} in Appendix \ref{ap:posterior}. Thanks to GAMA spectroscopy, we are able to compare our inferred redshift distribution with the spectroscopic redshifts. The gray contours in the left panel of Figure \ref{fig:gama_params_phot} show distributions of the spectroscopic redshifts and the stellar masses from the GAMA DR3 catalog. Our redshift estimates align well with the spectroscopic results, despite the photometric data being intrinsically less informative of redshift. 

In terms of stellar mass, we compare our stellar mass distribution to the distribution of the total formed stellar mass (\texttt{logmintsfh}) in the GAMA catalog. We simply combine the individual stellar masses in the GAMA catalog without considering the reported uncertainties. We find that our stellar mass distribution is on average 0.35 dex higher than that of GAMA. We note that the stellar mass in GAMA is derived for individual galaxies following the SPS model in \citet{Taylor2011}, where a $\tau$-model for SFH and the stellar evolution models in \citet{BC03} are used. Given these differences in SPS models between GAMA and our method, it is not surprising to see an offset between GAMA's stellar mass and ours. To account for this systematic effect, we add a constant 0.35 dex offset to all GAMA stellar masses and then compare with our results in Figures \ref{fig:gama_params_phot} and \ref{fig:gama_posterior}. We find that the GAMA stellar mass distribution is consistent with ours quite well, but is slightly skewed toward the lower-mass end. 
As with the mock test, constraints on metallicity and age are relatively less stringent. Additionally, the right panel in Figure \ref{fig:gama_params_phot} shows the distributions of the GAMA data and the synthetic photometric data from our inferred galaxy population. While a minor difference is observed in the $u-g$ color, the two distributions show excellent agreement in other bands.

\begin{figure}
 \centering
 \includegraphics[width=1\linewidth]{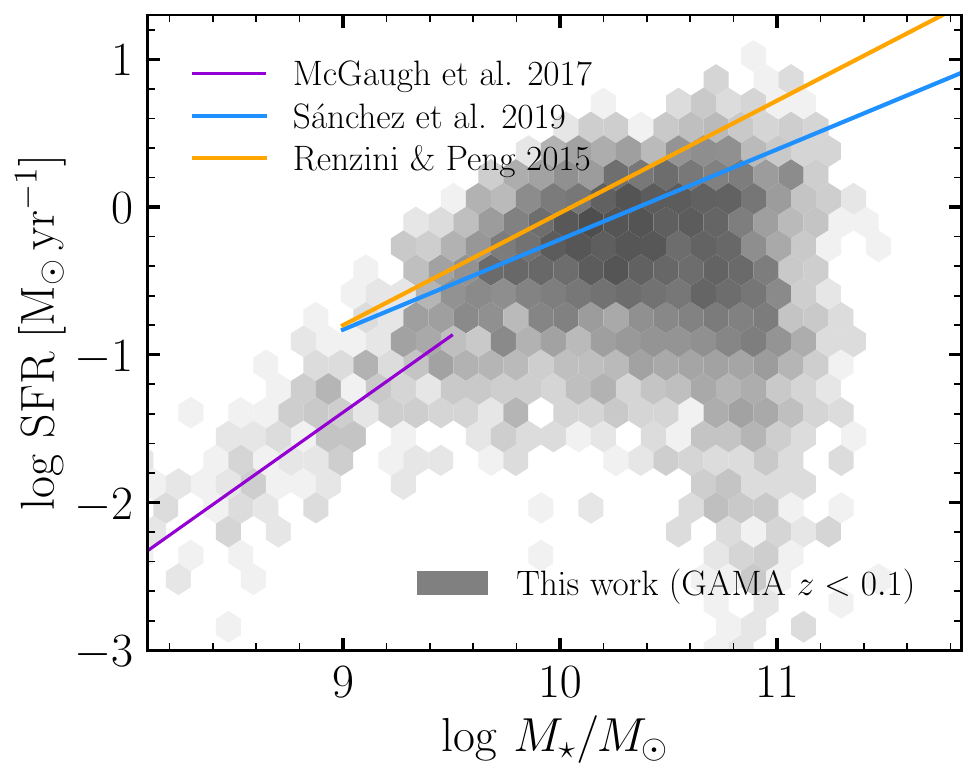}
 \caption{The distribution of the inferred galaxy population at $z<0.1$ on the $M_\star$--SFR plane. The SFMS and the quiescent population are clearly shown. The inferred galaxy population agrees well with the SFMSs from \citet{Renzini2015}, \citet{McGaugh2017}, and \citet{Sanchze2019}.
 \label{fig:gama_sfms} 
 }
\end{figure}

Leveraging the inferred population distribution, we can take slices and marginalize over irrelevant parameters to study correlations among key physical parameters. To showcase the capabilities of \popsed on such population-level analysis, we focus on constraining the SFMS for galaxies at $z<0.1$. With the samples drawn from the inferred population distribution in hand, we select samples with $z<0.1$ and calculate the average SFR within the past 0.1 Gyr and the surviving stellar mass using the fitting function in \citet{Hearin2021}\footnote{\url{https://github.com/ArgonneCPAC/dsps/blob/main/dsps/imf/surviving_mstar.py}}. Here we choose the surviving stellar mass rather than the total formed mass to better compare with literature results. The distribution of the inferred galaxy population on the $M_{\star}$--SFR plane is shown as the gray hexagons in Figure \ref{fig:gama_sfms}, colored on a logarithmic scale. Figure \ref{fig:gama_sfms} clearly reveals the main sequence of star-forming galaxies as well as a continuous transition from star-forming to quiescent at $10 < \log M_\star /M_\odot < 11.5$. 

To contextualize our findings, we further compare the derived galaxy distribution on the $M_\star$--SFR plane with literature results. \citet[][orange line in Figure \ref{fig:gama_sfms}]{Renzini2015} derived a linear SFMS using the SDSS galaxies at $0.02 < z < 0.085$, whose SFRs are estimated using H$\alpha$ following \citet{Brinchmann2004}. \citet[][blue line]{Sanchze2019} studied the SFMS for galaxies in the MaNGA survey \citep{Bundy2015} using the average SFR within the past 0.1~Gyr from SED fitting. Our results are qualitatively consistent with these results at $9.5 < \log M_\star/M_\odot < 11.5$. At the lower-mass end, \citet[][purple line]{McGaugh2017} explored the SFMS of low-surface-brightness galaxies at $D < 100$~Mpc using the H$\alpha$-based SFRs. Our results agree well with \citet{McGaugh2017} at $\log M_\star/M_\odot < 9.5$ and even capture the trend of the SFMS slope flattening with increasing stellar mass. 

We note that the positive correlation between $M_\star$ and SFR is also seen in the mock test (Figure \ref{fig:mock_params_phot}), where the SFH is designed to be independent of $M_\star$. However, the quiescent galaxy population and the shallower slope at low mass are nontrivial to capture and showcase the potential of \popsed for probing the dependence of SFH on $M_\star$. Nonetheless, it should be noted that proper weighting is needed to account for observational completeness when deriving the SFMS from the population distribution. We defer such analysis to future studies.

\section{Discussion}\label{sec:discuss}

\subsection{Advantage of Population-level Inference using \popsed}

Population-level analyses of galaxies provide important insights into key questions of astrophysics. However, the traditional ways of SED modeling on an individual galaxy basis are very expensive. Using traditional SED fitting methods, analyzing $10^5$ galaxies will take up to $2\times 10^6$ CPU hr. Even with the development of accelerated SED fitting \citep[e.g.,][]{Alsing2020ApJS, Hearin2021, Hahn2022sedflow, Khullar2022, Wang2023}, an analysis of $10^5$ galaxies will still take up to $\sim 10^{3}$ GPU hr. \popsed is able to recover the posterior of the population distribution for $\sim 10^{5}$ galaxies within $\sim 10$ GPU hr, 100 times faster than the SBI-based methods. In contrast to \popsed, it is challenging to change the SPS and noise models in SBI-based methods because of the cost of generating training data and retraining the SBI model.

Moreover, it is nontrivial to combine the posteriors of individual galaxies from SED fitting to construct a population-level distribution in a statistically rigorous way. Individual posteriors should not be multiplied directly because the prior distribution will be multiplied many times and dominate the resulting posterior. Hierarchical Bayesian models are often used to tackle this problem, and such population-level inference has been successfully applied to the studies of galaxy redshifts \citep{Leistedt2016,Malz2020,Alsing2022} and gravitational-wave sources \citep[e.g.,][]{Wong2020}. Taking the formulation in \citet{Malz2020}, the population posterior is often described by simple statistical models (e.g., Gaussian mixture models) parameterized by hyperparameters $\bm{\varphi}$. The posterior of the population distribution can be written as 

\begin{equation*}
    p(\bm{\varphi}|\{\bm{X}_i\}) = p(\bm{\varphi}) \cdot \Pi_{i=1}^{N} \int \frac{p(\bm{\theta}_i|\bm{X}_i) p(\bm{\theta}_i|\bm{\varphi})}{p(\bm{\theta}_i)} \md \bm{\theta}_i.
\end{equation*}
Evaluating this posterior requires the calculation of $N$ integrals, where the integrals are evaluated using Monte Carlo samples from individual posteriors $p(\bm{\theta}_i|\bm{X}_i)$. For each galaxy, one needs to save thousands of samples from its posterior to compute the integral. It is very expensive to store all the individual posteriors and computationally heavy to run MCMC chains to construct the population distribution in this fashion. It is also difficult to ensure that individual posteriors are accurate enough such that the combined population posterior is not biased when combining a large number of them.

\popsed bypasses the problem of deriving and combining individual posteriors, because we directly find the optimal solution for the population distribution by minimizing the distance between the observed data and the synthetic data. \popsed is much more efficient and robust at inferring the underlying population distribution. The ensemble of flows also provides an effective estimate of the posterior of the population distribution, shown as the light histograms in Figure \ref{fig:mock_params_phot} and \ref{fig:gama_params_phot}.

\subsection{Potential applications of \popsed}
Many science cases would benefit from having the population distribution in hand. As we demonstrate in \S\ref{sec:gama}, the population distribution can be marginalized to learn about the key relationships, such as the SFMS and stellar mass function. When applied to different galaxy samples that are selected differently (e.g., by colors), \popsed will be able to tell the underlying physical difference between the two galaxy populations. The \popsed framework can also be enhanced by integrating known spectroscopic redshifts, allowing it to concentrate primarily on inferring the physical properties rather than the redshift.

Accurately determining the redshift distribution of galaxies is crucial for extracting cosmological parameters from weak-lensing surveys \citep[e.g.,][]{Mandelbaum2018,Newman2022,Dalal2023}. In this work, we demonstrate that \popsed is capable of recovering the redshift distribution of GAMA galaxies very well. Such population-level analyses of galaxies \citep[e.g.,][]{Alsing2022} could greatly help the upcoming weak-lensing surveys including LSST \citep{LSST2019}, \textit{Euclid} \citep{Euclid2016}, and \textit{Roman} \citep{Spergel2015}

The \popsed{} framework can also be employed to generate synthetic observations in specified filters with different noise levels. These simulated observations can serve as a good reference for designing new surveys and assessing survey completeness \citep[e.g.,][]{Luo2023}. The target selection for dedicated studies (such as spectroscopic surveys) can now be done by selecting regions in the population distribution, and then translating them into the photometry (color) space. The framework also enables the exploration of outliers by sampling from low-probability regions in the population distribution \citep[e.g.,][]{Liang2023} and subsequently identifying objects in the photometric space that resemble these outliers.

Furthermore, the idea of population-level inference presented in this paper could be generalized to understand the populations behind galaxy spectra (e.g., DESI: \citealt{DESI-BGS}; PFS: \citealt{GreenePFS2022}), quasar spectra \citep[e.g.,][]{Sun2022}, and stellar spectra \citep[e.g., Gaia XP spectra: ][]{Zhang2023}.

\subsection{Limitations and Future Work}
\textbf{Runtime.}\ As the number of galaxies and bandpasses in photometric surveys increases, the bottleneck of running \popsed will be the cost of computing the Wasserstein distance. In this work, we use the Sinkhorn iteration to approximate the Wasserstein distance. However, the time complexity of the Sinkhorn algorithm is $O(MN)$, where $M$ and $N$ are the number of samples in the two discrete distributions \citep{Altschuler2017}. Thus, the cost of \popsed will grow faster with the number of galaxies than traditional methods. One possible solution is to use the ``sliced Wasserstein distance'' \citep[e.g.,][]{Bonneel2015,Kolouri2018}. Instead of solving optimal transport in high dimensions, one can slice the high-dimensional probability distributions into a number of one-dimensional distributions and calculate their Wasserstein distances. Therefore, the sliced Wasserstein distance is much faster to compute. We can also accelerate \popsed by calculating the Wasserstein distance using small batches, which also increases the stochasticity when training the normalizing flows. 

\textbf{Accuracy.}\ The SPS model in this work is emulated using a neural network, which is trained using synthetic spectra by sampling the parameter space. We notice that generating training samples and training such an emulator can be costly. The emulator might also perform poorly near the boundary of the prior used to train it. Furthermore, our SPS model does not include emission lines, making it inapplicable to narrowband photometric data. \citet{Hearin2021} presented \code{DSPS}, a fast and differentiable SPS model implemented in \code{jax} with no emulation. \code{DSPS} does not need training and can be directly used in \popsed once it is more developed and tested, as an alternative to the SPS emulator. 

\textbf{Robustness.}\ Because of the low constraining power of broadband photometry on the physical parameters, we take the annealing strategies (\S\ref{sec:train}) to make sure the normalizing flow smoothly converges from the uniform prior to the solution. A more informative prior distribution would likely make the training more stable. Such a prior might be learned from running \popsed on a small subset of the data. We also notice that the average SNR in the $u$ band is typically the lowest, thus it is specifically challenging for \popsed to extract information from the $u$-band data. A better treatment for the noise in the low SNR regime \citep[e.g.,][]{Lupton1999} in the forward model might help mitigate the inference bias in the $u-g$ color seen in Figure \ref{fig:gama_params_phot}.

\textbf{Selection effects.}\ The survey completeness and selection effects are not encoded in \popsed. One solution is to include a differentiable selection function in the forward model \citep[e.g.,][]{Alsing2022}. Nevertheless, if one has a well-defined survey completeness as a function of photometry, one can sample from the population distribution, calculate the corresponding completeness for each sample, and then reweight the samples to construct a new population distribution. We defer such studies to future works.

\section{Summary}
In this work, we propose \popsed, a novel framework to efficiently infer the distribution of physical parameters for an entire galaxy population using broadband photometric data. We focus on the \textit{population-level analysis} of galaxies, rather than inferring physical properties for individual objects. We successfully applied \popsed to GAMA DR3 data and obtained a posterior of the population distribution that is consistent with spectroscopic results. Our main findings and prospects are as follows:

\begin{itemize}
    \item The overall structure of \popsed is highlighted in Figure \ref{fig:flow}. We approximate the population-level distribution of physical parameters using a normalizing flow (\S\ref{sec:nde}), a highly flexible neural density estimator. We then forward model the synthetic photometry of galaxies using the samples generated from the normalizing flow and the emulator for the SPS model (\S\ref{sec:spsmodel} and \S\ref{sec:emulator}). The normalizing flow is trained to minimize the Wasserstein distance (\S\ref{sec:wasserstein}) between this synthetic photometric data and the observations. After training, the flow is able to approximate the population posterior and imitate the observed photometric data distribution.
    
    \item \popsed is able to analyze $10^{5}$ galaxies within 1 GPU hr, being $\sim 10^{5}$ times faster than traditional MCMC methods and $\sim 100$ times faster than SBI-based methods. It also circumvents the problem of combining individual posteriors to derive the population distribution.
    
    \item We validate \popsed through a carefully designed mock observation (\S\ref{sec:mock}). The results prove that \popsed can accurately recover the underlying population posterior, especially for key parameters such as stellar mass, redshift, and SFR (Figure \ref{fig:mock_params_phot}).
    
    \item We then apply our method to 83,692 galaxies from the GAMA survey (\S\ref{sec:gama}). Our inferred population distribution shows a remarkable agreement with the distribution of spectroscopic redshift and stellar mass, despite photometric data being less informative for constraining redshift (Figure \ref{fig:gama_params_phot}). We further demonstrate the versatility of \popsed by studying the SFMS for galaxies at $z<0.1$. Our results are in good agreement with the literature results (Figure \ref{fig:gama_sfms}). 

    \item \popsed holds promise for application to future photometric surveys to derive the redshift distribution for weak-lensing studies and understand galaxy formation and evolution. The idea of population-level inference can also be applied to analyze stellar spectra and galaxy spectra. 
\end{itemize}


\section*{Acknowledgment}
We thank the anonymous reviewer for useful comments that make the manuscript much clearer. J.L. is grateful for discussions with Kaixuan Huang, Jenny Greene, Sihao Cheng, Yuan-Sen Ting, He Jia, Alexie Leauthaud, Rachel Mandelbaum, Meng Gu, Yifei Luo, Andy Goulding, Alexa Villaume, and Runquan Guan. We are grateful for the comments from anonymous reviewers of the 2023 ICML ML4astro workshop. 

This work was supported by the AI Accelerator program of the Schmidt Futures Foundation. The authors are pleased to acknowledge that the work reported on in this paper was substantially performed using the Princeton Research Computing resources at Princeton University, which is a consortium of groups led by the Princeton Institute for Computational Science and Engineering (PICSciE) and the Office of Information Technology's Research Computing.

GAMA is a joint European--Australasian project based around a spectroscopic campaign using the Anglo-Australian Telescope. The GAMA input catalog is based on data taken from the Sloan Digital Sky Survey and the UKIRT Infrared Deep Sky Survey. Complementary imaging of the GAMA regions is being obtained by a number of independent survey programs, including GALEX MIS, VST KiDS, VISTA VIKING, WISE, Herschel-ATLAS, GMRT, and ASKAP, providing UV to radio coverage. GAMA is funded by the STFC (UK), the ARC (Australia), the AAO, and the participating institutions. The GAMA website is http://www.gama-survey.org/.

\software{\href{http://www.numpy.org}{\code{NumPy}} \citep{Numpy},
          \href{https://www.astropy.org/}{\code{Astropy}} \citep{astropy,astropy2018,astropy2022},  \href{https://www.scipy.org}{\code{SciPy}} \citep{scipy}, \href{https://matplotlib.org}{\code{Matplotlib}} \citep{matplotlib},
          \href{https://pytorch.org/}{\code{PyTorch}} \citep{pytorch},
          \href{https://github.com/justinalsing/speculator}{\code{Speculator}} \citep{Alsing2020ApJS},
          \href{https://github.com/cconroy20/fsps}{\code{FSPS}} \citep{Conroy2009,Conroy2010},
          \href{https://dfm.io/python-fsps/current/}{\code{python-fsps}} \citep{pyfsps},
          \href{https://github.com/bd-j/sedpy}{\code{sedpy}} \citep{sedpy}, \href{https://github.com/desihub/speclite}{\code{speclite}},
          \href{https://github.com/mackelab/sbi}{\code{sbi}} \citep{SBI},
          \href{https://www.kernel-operations.io/geomloss/}{\code{geomloss}} \citep{geomloss}, \href{https://www.kernel-operations.io/keops/index.html}{\code{pyKeOps}} \citep{pykeops}.
          }

\appendix

\section{Details on Training the Spectrum Emulator}\label{ap:sps}

Our spectrum emulator is very similar to the emulators in \citet{Alsing2020ApJS} and \citet{Hahn2022sedflow}. To train the emulator, we generate \textit{rest-frame} spectra $L_\lambda(\bm{\theta})$ from 1000\AA{} to 60,000\AA{} for galaxies with fixed stellar masses $M_\star = 1\ M_\odot$. We sample the parameter space according to the prior distributions listed in Table \ref{tab:parameters} to make the training data more representative. Uninformative priors are employed to avoid introducing any bias when training the emulator. The SFH coefficients $\beta_i$ are sampled from a flat Dirichlet prior, which is equivalent to a uniform distribution over the open standard three-dimensional simplex. Following \citet{Betancourt2012}, we first sample $\kappa_j\sim \mathrm{Uniform}\,(0, 1),\ j={1, 2, 3}$, then transform $\kappa_j$ to $\beta_i$. Uniform priors are used for other SPS parameters. When generating the training spectra, the redshift $z$ is only used to calculate the galaxy age $t_{\rm age}$ which marks the endpoint of the SFH. In the end, we generate $3\times 10^6$ rest-frame spectra to train the emulator and $10^4$ spectra for validation.

\begin{figure*}
    \centering
    \vbox{ 
		\centering
		\includegraphics[width=1\linewidth]{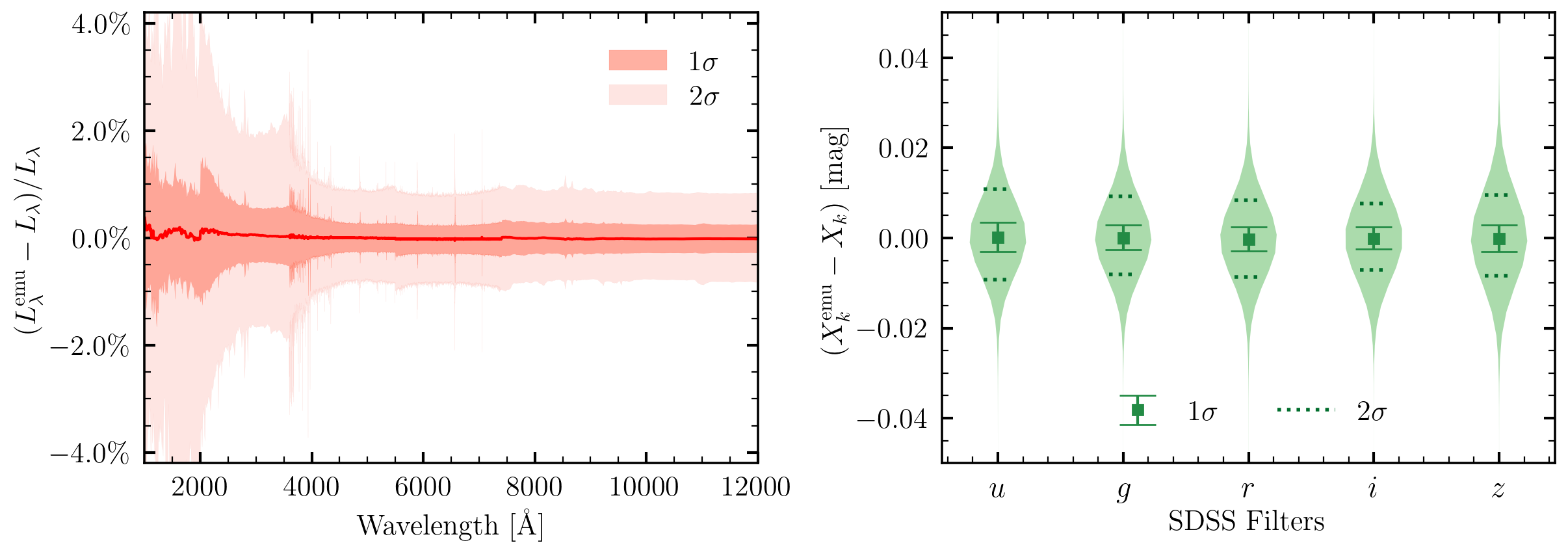}
	}
    \caption{Performance of the galaxy spectrum emulator. The left panel shows the fractional error of the emulated rest-frame spectrum $L_{\lambda}$ at 1000--12000 \AA{}. The shaded regions correspond to the 68th ($1\sigma$) and 95th ($2\sigma$) percentiles of the fractional error distribution. The right panel shows the emulation error of the broadband photometry $X_k$ in the SDSS $ugriz$ bands for galaxies at $z=0$. Here we use the SDSS $ugriz$ filters as a demonstration and take the transmission curves from \citet{Doi2010}, assuming an airmass of 1.3. The emulator achieves a high accuracy of $<1\%$ for $\lambda > 4000$\AA, but suffers at the very blue end. However, the emulated photometry is accurate to $<0.02$ mag, which is negligible for analyzing nearby galaxies in large photometric surveys. 
    }
    \label{fig:emulator_accuracy}
\end{figure*}

We split the full wavelength range into five bins: 1000--2000\AA, 2000--3600\AA, 3600--5500\AA, 5500--7410\AA, 7410--60000\AA, and we train one emulator for each wavelength bin separately. In order to reduce the dimensionality, the rest-frame spectra are first compressed using the principal component analysis (PCA) technique \citep{Alsing2020ApJS}. We use $N_{\rm bases} = 80, 50, 50, 50, 50$ PCA bases for each wavelength bin, and we find these PCA bases are sufficient to recover the training spectra to a high accuracy ($1\sigma$ error $<0.2\%$). Then we train a five-layer neural network to predict the PCA coefficients given the SPS parameters $\bm{\theta}$. We adopt the customized activation function that is used in \citet{Alsing2020ApJS}: $a(x) = \gamma x + (1 - \gamma)\, x \cdot \mathrm{sigmoid}(\beta x)$, where $\beta$ and $\gamma$ are trainable parameters. This activation function combines the Swish function (also known as SiLU; \citealt{Elfwing2017,Ramachandran2017}) with a linear component and works better than the commonly used ones, such as Sigmoid or ReLU, on emulating galaxy spectra. The neural network is implemented in \code{PyTorch} \citep{pytorch} and trained using the \code{Adam} optimizer \citep{Adam}, similar to \citet{Alsing2020ApJS}.

Finally, we calculate the observed spectrum of a galaxy with a stellar mass $M_\star$ at redshift $z$ by red-shifting and scaling the predicted rest-frame spectrum $L_{\lambda}(\bm{\theta})$, following 
\begin{equation}\label{eq:redshift}
    l_\lambda(\bm{\theta}) = \frac{L_{\lambda / (1+z)}(\bm{\theta})\cdot (M_\star/M_\odot)}{4\pi d_{L}^2(z) (1+z)},
\end{equation}
where $d_L(z)$ is the luminosity distance. We further convolve $l_\lambda(\bm{\theta})$ with the transmission curves $R_k(\lambda)$ of the SDSS filters in \citet{Doi2010} to obtain the noiseless flux $f_k$ in the SDSS bands\footnote{The transmission curves are calculated as the reference response of the filters multiplied by the atmospheric transmission at an airmass 1.3. We note that this set of transmission curves is different from the one measured by James E. Gunn in 2001 (\url{https://www.sdss4.org/instruments/camera/}). The $u$-band transmission curve has changed most significantly from 2001 to 2010.}:
\begin{equation}\label{eq:convolve}
    f_k(\bm{\theta}) = \int l_\lambda(\bm{\theta}) R_k(\lambda) \md \lambda,\quad k = \{u, g, r, i, z\}.
\end{equation}
We note that these operations are not embedded in the neural networks, but are rather post-processing.

We show the validation accuracy of the spectrum emulator $l_{\lambda}(\bm{\theta})$ at $z=0$ in Figure \ref{fig:emulator_accuracy}, where we limit our wavelength range to 1000--12000\AA{} for brevity. The emulator achieves $<1\%$ accuracy at $3000\,\rm{\AA} < \lambda < 12000\,\rm{\AA}$ and $\sim 2\%$ accuracy at $1000\,\rm{\AA} < \lambda < 3000\,\rm{\AA}$. The errors on the predicted broadband SEDs for galaxies at $z=0$ are shown in the right panel of Figure \ref{fig:emulator_accuracy}. The emulator is able to predict the broadband photometry to an accuracy of $\sim 0.01$ mag. This accuracy is more than enough for the population-level analysis of galaxy SEDs in photometric surveys. Since we are mostly interested in low- to intermediate-redshift galaxies ($z<1$) in this work, the emulation error at the very blue end does not affect the inference, since the observation noise almost always dominates the photometry error. However, if one wants to study higher-redshift galaxies, a better emulator at the blue end is required.

\section{Full population posteriors}\label{ap:posterior}
\begin{figure*}
 \centering
 \includegraphics[width=1\linewidth]{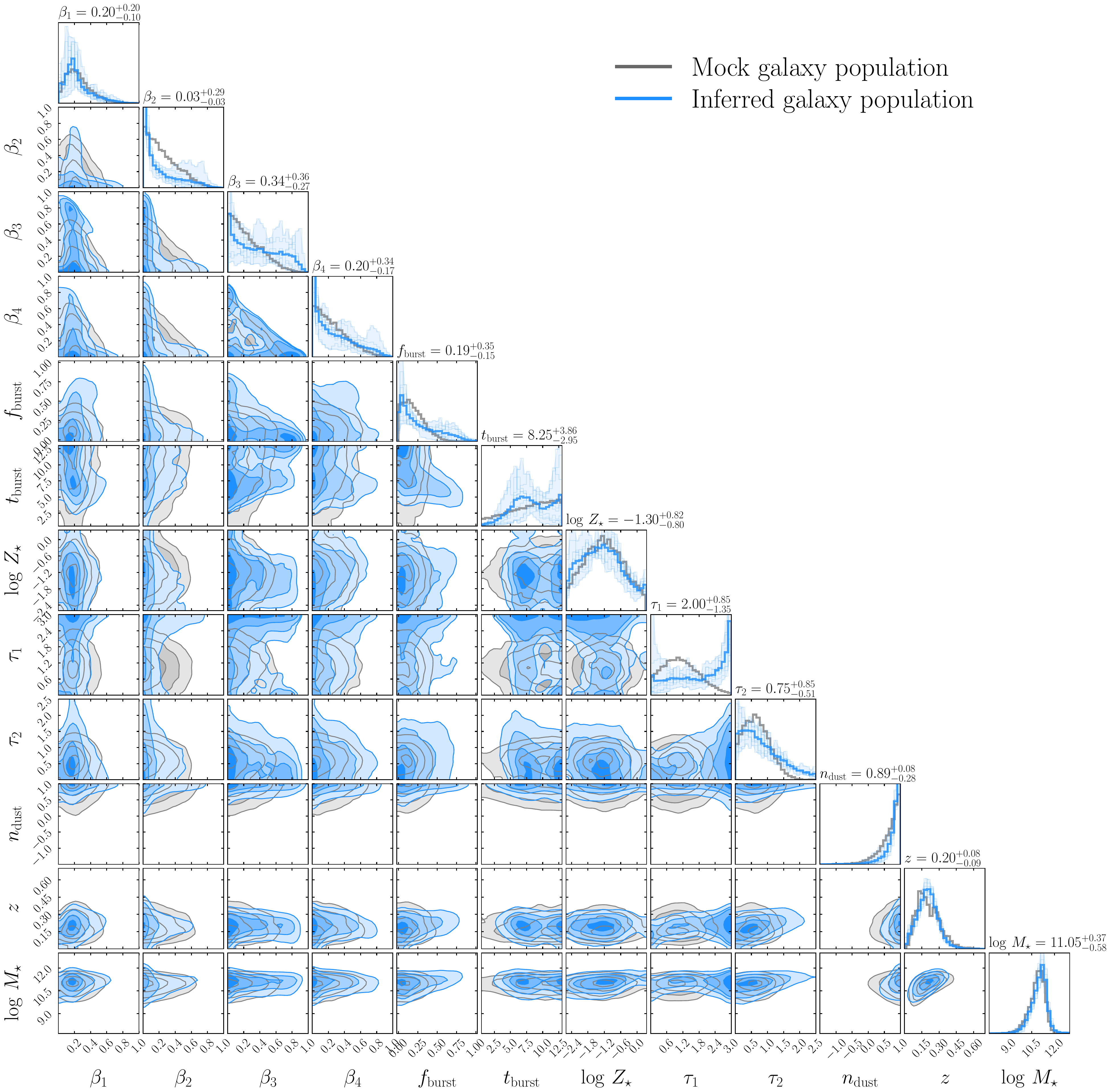}
 \caption{Similar to the left panel in Figure \ref{fig:mock_params_phot}, but showing the distributions of the mock galaxy population (gray) and the inferred galaxy population (blue) in the full parameter space. 
 \label{fig:mock_posterior} 
 }
\end{figure*}

\begin{figure*}
 \centering
 \includegraphics[width=1\linewidth]{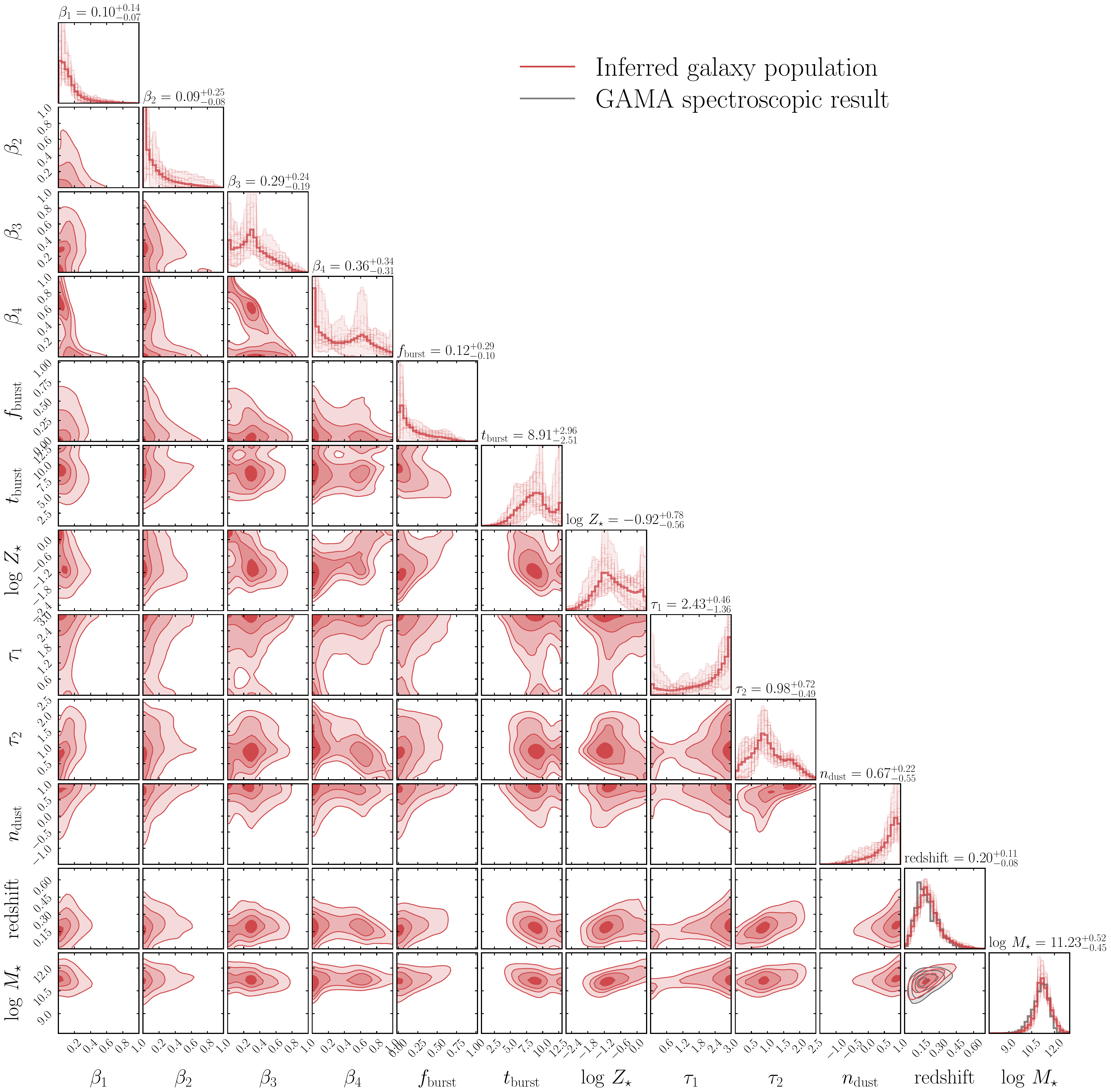}
 \caption{Similar to the left panel in Figure \ref{fig:gama_params_phot}, but showing the distributions of the inferred galaxy population (red) in the full parameter space. For comparison, the gray contours show the joint distribution of the spectroscopic redshift and the stellar mass from the GAMA catalog.
 \label{fig:gama_posterior}
 }
\end{figure*}

Here we present the full population posteriors of galaxies from the mock galaxy population (Figure \ref{fig:mock_posterior}) and the GAMA sample (Figure \ref{fig:gama_posterior}). 

\bibliography{citation}{}

\begin{thebibliography}{}
\expandafter\ifx\csname natexlab\endcsname\relax\def\natexlab#1{#1}\fi
\providecommand{\url}[1]{\href{#1}{#1}}
\providecommand{\dodoi}[1]{doi:~\href{http://doi.org/#1}{\nolinkurl{#1}}}
\providecommand{\doeprint}[1]{\href{http://ascl.net/#1}{\nolinkurl{http://ascl.net/#1}}}
\providecommand{\doarXiv}[1]{\href{https://arxiv.org/abs/#1}{\nolinkurl{https://arxiv.org/abs/#1}}}

\bibitem[{{Abazajian} {et~al.}(2009){Abazajian}, {Adelman-McCarthy},
  {Ag{\"u}eros}, {Allam}, {Allende Prieto}, {An}, {Anderson}, {Anderson},
  {Annis}, {Bahcall}, {Bailer-Jones}, {Barentine}, {Bassett}, {Becker},
  {Beers}, {Bell}, {Belokurov}, {Berlind}, {Berman}, {Bernardi}, {Bickerton},
  {Bizyaev}, {Blakeslee}, {Blanton}, {Bochanski}, {Boroski}, {Brewington},
  {Brinchmann}, {Brinkmann}, {Brunner}, {Budav{\'a}ri}, {Carey}, {Carliles},
  {Carr}, {Castander}, {Cinabro}, {Connolly}, {Csabai}, {Cunha}, {Czarapata},
  {Davenport}, {de Haas}, {Dilday}, {Doi}, {Eisenstein}, {Evans}, {Evans},
  {Fan}, {Friedman}, {Frieman}, {Fukugita}, {G{\"a}nsicke}, {Gates},
  {Gillespie}, {Gilmore}, {Gonzalez}, {Gonzalez}, {Grebel}, {Gunn},
  {Gy{\"o}ry}, {Hall}, {Harding}, {Harris}, {Harvanek}, {Hawley}, {Hayes},
  {Heckman}, {Hendry}, {Hennessy}, {Hindsley}, {Hoblitt}, {Hogan}, {Hogg},
  {Holtzman}, {Hyde}, {Ichikawa}, {Ichikawa}, {Im}, {Ivezi{\'c}}, {Jester},
  {Jiang}, {Johnson}, {Jorgensen}, {Juri{\'c}}, {Kent}, {Kessler}, {Kleinman},
  {Knapp}, {Konishi}, {Kron}, {Krzesinski}, {Kuropatkin}, {Lampeitl},
  {Lebedeva}, {Lee}, {Lee}, {French Leger}, {L{\'e}pine}, {Li}, {Lima}, {Lin},
  {Long}, {Loomis}, {Loveday}, {Lupton}, {Magnier}, {Malanushenko},
  {Malanushenko}, {Mandelbaum}, {Margon}, {Marriner}, {Mart{\'\i}nez-Delgado},
  {Matsubara}, {McGehee}, {McKay}, {Meiksin}, {Morrison}, {Mullally}, {Munn},
  {Murphy}, {Nash}, {Nebot}, {Neilsen}, {Newberg}, {Newman}, {Nichol},
  {Nicinski}, {Nieto-Santisteban}, {Nitta}, {Okamura}, {Oravetz}, {Ostriker},
  {Owen}, {Padmanabhan}, {Pan}, {Park}, {Pauls}, {Peoples}, {Percival}, {Pier},
  {Pope}, {Pourbaix}, {Price}, {Purger}, {Quinn}, {Raddick}, {Re Fiorentin},
  {Richards}, {Richmond}, {Riess}, {Rix}, {Rockosi}, {Sako}, {Schlegel},
  {Schneider}, {Scholz}, {Schreiber}, {Schwope}, {Seljak}, {Sesar}, {Sheldon},
  {Shimasaku}, {Sibley}, {Simmons}, {Sivarani}, {Allyn Smith}, {Smith},
  {Smol{\v{c}}i{\'c}}, {Snedden}, {Stebbins}, {Steinmetz}, {Stoughton},
  {Strauss}, {SubbaRao}, {Suto}, {Szalay}, {Szapudi}, {Szkody}, {Tanaka},
  {Tegmark}, {Teodoro}, {Thakar}, {Tremonti}, {Tucker}, {Uomoto}, {Vanden
  Berk}, {Vandenberg}, {Vidrih}, {Vogeley}, {Voges}, {Vogt}, {Wadadekar},
  {Watters}, {Weinberg}, {West}, {White}, {Wilhite}, {Wonders}, {Yanny},
  {Yocum}, {York}, {Zehavi}, {Zibetti}, \& {Zucker}}]{SDSS-DR7}
{Abazajian}, K.~N., {Adelman-McCarthy}, J.~K., {Ag{\"u}eros}, M.~A., {et~al.}
  2009, \apjs, 182, 543, \dodoi{10.1088/0067-0049/182/2/543}

\bibitem[{{Alsing} {et~al.}(2019){Alsing}, {Charnock}, {Feeney}, \&
  {Wandelt}}]{Alsing2019}
{Alsing}, J., {Charnock}, T., {Feeney}, S., \& {Wandelt}, B. 2019, \mnras, 488,
  4440, \dodoi{10.1093/mnras/stz1960}

\bibitem[{{Alsing} {et~al.}(2022){Alsing}, {Peiris}, {Mortlock}, {Leja}, \&
  {Leistedt}}]{Alsing2022}
{Alsing}, J., {Peiris}, H., {Mortlock}, D., {Leja}, J., \& {Leistedt}, B. 2022,
  arXiv e-prints, arXiv:2207.05819.
\newblock \doarXiv{2207.05819}

\bibitem[{{Alsing} {et~al.}(2020){Alsing}, {Peiris}, {Leja}, {Hahn}, {Tojeiro},
  {Mortlock}, {Leistedt}, {Johnson}, \& {Conroy}}]{Alsing2020ApJS}
{Alsing}, J., {Peiris}, H., {Leja}, J., {et~al.} 2020, \apjs, 249, 5,
  \dodoi{10.3847/1538-4365/ab917f}

\bibitem[{Altschuler {et~al.}(2017)Altschuler, Niles-Weed, \&
  Rigollet}]{Altschuler2017}
Altschuler, J., Niles-Weed, J., \& Rigollet, P. 2017, in Advances in Neural
  Information Processing Systems, ed. I.~Guyon, U.~V. Luxburg, S.~Bengio,
  H.~Wallach, R.~Fergus, S.~Vishwanathan, \& R.~Garnett, Vol.~30 (Curran
  Associates, Inc.).
\newblock
  \url{https://proceedings.neurips.cc/paper_files/paper/2017/file/491442df5f88c6aa018e86dac21d3606-Paper.pdf}

\bibitem[{{Ambrogioni} {et~al.}(2018){Ambrogioni}, {G{\"u}{\c{c}}l{\"u}},
  {G{\"u}{\c{c}}l{\"u}t{\"u}rk}, {Hinne}, {Maris}, \& {van
  Gerven}}]{Ambrogioni2018}
{Ambrogioni}, L., {G{\"u}{\c{c}}l{\"u}}, U., {G{\"u}{\c{c}}l{\"u}t{\"u}rk}, Y.,
  {et~al.} 2018, arXiv e-prints, arXiv:1805.11284,
  \dodoi{10.48550/arXiv.1805.11284}

\bibitem[{{Arjovsky} {et~al.}(2017){Arjovsky}, {Chintala}, \&
  {Bottou}}]{Arjovsky2017}
{Arjovsky}, M., {Chintala}, S., \& {Bottou}, L. 2017, arXiv e-prints,
  arXiv:1701.07875, \dodoi{10.48550/arXiv.1701.07875}

\bibitem[{{Astropy Collaboration} {et~al.}(2013){Astropy Collaboration},
  {Robitaille}, {Tollerud}, {Greenfield}, {Droettboom}, {Bray}, {Aldcroft},
  {Davis}, {Ginsburg}, {Price-Whelan}, {Kerzendorf}, {Conley}, {Crighton},
  {Barbary}, {Muna}, {Ferguson}, {Grollier}, {Parikh}, {Nair}, {Unther},
  {Deil}, {Woillez}, {Conseil}, {Kramer}, {Turner}, {Singer}, {Fox}, {Weaver},
  {Zabalza}, {Edwards}, {Azalee Bostroem}, {Burke}, {Casey}, {Crawford},
  {Dencheva}, {Ely}, {Jenness}, {Labrie}, {Lim}, {Pierfederici}, {Pontzen},
  {Ptak}, {Refsdal}, {Servillat}, \& {Streicher}}]{astropy}
{Astropy Collaboration}, {Robitaille}, T.~P., {Tollerud}, E.~J., {et~al.} 2013,
  \aap, 558, A33, \dodoi{10.1051/0004-6361/201322068}

\bibitem[{{Astropy Collaboration} {et~al.}(2018){Astropy Collaboration},
  {Price-Whelan}, {Sip{\H{o}}cz}, {G{\"u}nther}, {Lim}, {Crawford}, {Conseil},
  {Shupe}, {Craig}, {Dencheva}, {Ginsburg}, {VanderPlas}, {Bradley},
  {P{\'e}rez-Su{\'a}rez}, {de Val-Borro}, {Aldcroft}, {Cruz}, {Robitaille},
  {Tollerud}, {Ardelean}, {Babej}, {Bach}, {Bachetti}, {Bakanov}, {Bamford},
  {Barentsen}, {Barmby}, {Baumbach}, {Berry}, {Biscani}, {Boquien}, {Bostroem},
  {Bouma}, {Brammer}, {Bray}, {Breytenbach}, {Buddelmeijer}, {Burke},
  {Calderone}, {Cano Rodr{\'\i}guez}, {Cara}, {Cardoso}, {Cheedella}, {Copin},
  {Corrales}, {Crichton}, {D'Avella}, {Deil}, {Depagne}, {Dietrich}, {Donath},
  {Droettboom}, {Earl}, {Erben}, {Fabbro}, {Ferreira}, {Finethy}, {Fox},
  {Garrison}, {Gibbons}, {Goldstein}, {Gommers}, {Greco}, {Greenfield},
  {Groener}, {Grollier}, {Hagen}, {Hirst}, {Homeier}, {Horton}, {Hosseinzadeh},
  {Hu}, {Hunkeler}, {Ivezi{\'c}}, {Jain}, {Jenness}, {Kanarek}, {Kendrew},
  {Kern}, {Kerzendorf}, {Khvalko}, {King}, {Kirkby}, {Kulkarni}, {Kumar},
  {Lee}, {Lenz}, {Littlefair}, {Ma}, {Macleod}, {Mastropietro}, {McCully},
  {Montagnac}, {Morris}, {Mueller}, {Mumford}, {Muna}, {Murphy}, {Nelson},
  {Nguyen}, {Ninan}, {N{\"o}the}, {Ogaz}, {Oh}, {Parejko}, {Parley}, {Pascual},
  {Patil}, {Patil}, {Plunkett}, {Prochaska}, {Rastogi}, {Reddy Janga},
  {Sabater}, {Sakurikar}, {Seifert}, {Sherbert}, {Sherwood-Taylor}, {Shih},
  {Sick}, {Silbiger}, {Singanamalla}, {Singer}, {Sladen}, {Sooley},
  {Sornarajah}, {Streicher}, {Teuben}, {Thomas}, {Tremblay}, {Turner},
  {Terr{\'o}n}, {van Kerkwijk}, {de la Vega}, {Watkins}, {Weaver}, {Whitmore},
  {Woillez}, {Zabalza}, \& {Astropy Contributors}}]{astropy2018}
{Astropy Collaboration}, {Price-Whelan}, A.~M., {Sip{\H{o}}cz}, B.~M., {et~al.}
  2018, \aj, 156, 123, \dodoi{10.3847/1538-3881/aabc4f}

\bibitem[{{Astropy Collaboration} {et~al.}(2022){Astropy Collaboration},
  {Price-Whelan}, {Lim}, {Earl}, {Starkman}, {Bradley}, {Shupe}, {Patil},
  {Corrales}, {Brasseur}, {N{\"o}the}, {Donath}, {Tollerud}, {Morris},
  {Ginsburg}, {Vaher}, {Weaver}, {Tocknell}, {Jamieson}, {van Kerkwijk},
  {Robitaille}, {Merry}, {Bachetti}, {G{\"u}nther}, {Aldcroft},
  {Alvarado-Montes}, {Archibald}, {B{\'o}di}, {Bapat}, {Barentsen},
  {Baz{\'a}n}, {Biswas}, {Boquien}, {Burke}, {Cara}, {Cara}, {Conroy},
  {Conseil}, {Craig}, {Cross}, {Cruz}, {D'Eugenio}, {Dencheva}, {Devillepoix},
  {Dietrich}, {Eigenbrot}, {Erben}, {Ferreira}, {Foreman-Mackey}, {Fox},
  {Freij}, {Garg}, {Geda}, {Glattly}, {Gondhalekar}, {Gordon}, {Grant},
  {Greenfield}, {Groener}, {Guest}, {Gurovich}, {Handberg}, {Hart},
  {Hatfield-Dodds}, {Homeier}, {Hosseinzadeh}, {Jenness}, {Jones}, {Joseph},
  {Kalmbach}, {Karamehmetoglu}, {Ka{\l}uszy{\'n}ski}, {Kelley}, {Kern},
  {Kerzendorf}, {Koch}, {Kulumani}, {Lee}, {Ly}, {Ma}, {MacBride}, {Maljaars},
  {Muna}, {Murphy}, {Norman}, {O'Steen}, {Oman}, {Pacifici}, {Pascual},
  {Pascual-Granado}, {Patil}, {Perren}, {Pickering}, {Rastogi}, {Roulston},
  {Ryan}, {Rykoff}, {Sabater}, {Sakurikar}, {Salgado}, {Sanghi}, {Saunders},
  {Savchenko}, {Schwardt}, {Seifert-Eckert}, {Shih}, {Jain}, {Shukla}, {Sick},
  {Simpson}, {Singanamalla}, {Singer}, {Singhal}, {Sinha}, {Sip{\H{o}}cz},
  {Spitler}, {Stansby}, {Streicher}, {{\v{S}}umak}, {Swinbank}, {Taranu},
  {Tewary}, {Tremblay}, {de Val-Borro}, {Van Kooten}, {Vasovi{\'c}}, {Verma},
  {de Miranda Cardoso}, {Williams}, {Wilson}, {Winkel}, {Wood-Vasey}, {Xue},
  {Yoachim}, {Zhang}, {Zonca}, \& {Astropy Project Contributors}}]{astropy2022}
{Astropy Collaboration}, {Price-Whelan}, A.~M., {Lim}, P.~L., {et~al.} 2022,
  \apj, 935, 167, \dodoi{10.3847/1538-4357/ac7c74}

\bibitem[{{Baldry} {et~al.}(2018){Baldry}, {Liske}, {Brown}, {Robotham},
  {Driver}, {Dunne}, {Alpaslan}, {Brough}, {Cluver}, {Eardley}, {Farrow},
  {Heymans}, {Hildebrandt}, {Hopkins}, {Kelvin}, {Loveday}, {Moffett},
  {Norberg}, {Owers}, {Taylor}, {Wright}, {Bamford}, {Bland-Hawthorn},
  {Bourne}, {Bremer}, {Colless}, {Conselice}, {Croom}, {Davies}, {Foster},
  {Grootes}, {Holwerda}, {Jones}, {Kafle}, {Kuijken}, {Lara-Lopez},
  {L{\'o}pez-S{\'a}nchez}, {Meyer}, {Phillipps}, {Sutherland}, {van Kampen}, \&
  {Wilkins}}]{Baldry2018}
{Baldry}, I.~K., {Liske}, J., {Brown}, M.~J.~I., {et~al.} 2018, \mnras, 474,
  3875, \dodoi{10.1093/mnras/stx3042}

\bibitem[{{Bernton} {et~al.}(2019){Bernton}, {Jacob}, {Gerber}, \&
  {Robert}}]{Bernton2019}
{Bernton}, E., {Jacob}, P.~E., {Gerber}, M., \& {Robert}, C.~P. 2019, arXiv
  e-prints, arXiv:1905.03747, \dodoi{10.48550/arXiv.1905.03747}

\bibitem[{{Betancourt}(2012)}]{Betancourt2012}
{Betancourt}, M. 2012, in American Institute of Physics Conference Series, Vol.
  1443, Bayesian Inference and Maximum Entropy Methods in Science and
  Engineering: 31st International Workshop on Bayesian Inference and Maximum
  Entropy Methods in Science and Engineering, ed. P.~{Goyal}, A.~{Giffin},
  K.~H. {Knuth}, \& E.~{Vrscay}, 157--164, \dodoi{10.1063/1.3703631}

\bibitem[{Bonneel {et~al.}(2015)Bonneel, Rabin, Peyr{\'e}, \&
  Pfister}]{Bonneel2015}
Bonneel, N., Rabin, J., Peyr{\'e}, G., \& Pfister, H. 2015, Journal of
  Mathematical Imaging and Vision, 51, 22

\bibitem[{{Bovy} {et~al.}(2011){Bovy}, {Hogg}, \& {Roweis}}]{Bovy2011}
{Bovy}, J., {Hogg}, D.~W., \& {Roweis}, S.~T. 2011, Annals of Applied
  Statistics, 5, 1657, \dodoi{10.1214/10-AOAS439}

\bibitem[{{Brinchmann} {et~al.}(2004){Brinchmann}, {Charlot}, {White},
  {Tremonti}, {Kauffmann}, {Heckman}, \& {Brinkmann}}]{Brinchmann2004}
{Brinchmann}, J., {Charlot}, S., {White}, S.~D.~M., {et~al.} 2004, \mnras, 351,
  1151, \dodoi{10.1111/j.1365-2966.2004.07881.x}

\bibitem[{{Bruzual} \& {Charlot}(2003)}]{BC03}
{Bruzual}, G., \& {Charlot}, S. 2003, \mnras, 344, 1000,
  \dodoi{10.1046/j.1365-8711.2003.06897.x}

\bibitem[{{Bundy} {et~al.}(2015){Bundy}, {Bershady}, {Law}, {Yan}, {Drory},
  {MacDonald}, {Wake}, {Cherinka}, {S{\'a}nchez-Gallego}, {Weijmans}, {Thomas},
  {Tremonti}, {Masters}, {Coccato}, {Diamond-Stanic}, {Arag{\'o}n-Salamanca},
  {Avila-Reese}, {Badenes}, {Falc{\'o}n-Barroso}, {Belfiore}, {Bizyaev},
  {Blanc}, {Bland-Hawthorn}, {Blanton}, {Brownstein}, {Byler}, {Cappellari},
  {Conroy}, {Dutton}, {Emsellem}, {Etherington}, {Frinchaboy}, {Fu}, {Gunn},
  {Harding}, {Johnston}, {Kauffmann}, {Kinemuchi}, {Klaene}, {Knapen},
  {Leauthaud}, {Li}, {Lin}, {Maiolino}, {Malanushenko}, {Malanushenko}, {Mao},
  {Maraston}, {McDermid}, {Merrifield}, {Nichol}, {Oravetz}, {Pan}, {Parejko},
  {Sanchez}, {Schlegel}, {Simmons}, {Steele}, {Steinmetz}, {Thanjavur},
  {Thompson}, {Tinker}, {van den Bosch}, {Westfall}, {Wilkinson}, {Wright},
  {Xiao}, \& {Zhang}}]{Bundy2015}
{Bundy}, K., {Bershady}, M.~A., {Law}, D.~R., {et~al.} 2015, \apj, 798, 7,
  \dodoi{10.1088/0004-637X/798/1/7}

\bibitem[{{Calzetti} {et~al.}(2000){Calzetti}, {Armus}, {Bohlin}, {Kinney},
  {Koornneef}, \& {Storchi-Bergmann}}]{Calzetti2000}
{Calzetti}, D., {Armus}, L., {Bohlin}, R.~C., {et~al.} 2000, \apj, 533, 682,
  \dodoi{10.1086/308692}

\bibitem[{{Carnall} {et~al.}(2019){Carnall}, {Leja}, {Johnson}, {McLure},
  {Dunlop}, \& {Conroy}}]{Carnall2019}
{Carnall}, A.~C., {Leja}, J., {Johnson}, B.~D., {et~al.} 2019, \apj, 873, 44,
  \dodoi{10.3847/1538-4357/ab04a2}

\bibitem[{{Carnall} {et~al.}(2018){Carnall}, {McLure}, {Dunlop}, \&
  {Dav{\'e}}}]{Carnall2018}
{Carnall}, A.~C., {McLure}, R.~J., {Dunlop}, J.~S., \& {Dav{\'e}}, R. 2018,
  \mnras, 480, 4379, \dodoi{10.1093/mnras/sty2169}

\bibitem[{{Chabrier}(2003)}]{Chabrier2003}
{Chabrier}, G. 2003, \pasp, 115, 763, \dodoi{10.1086/376392}

\bibitem[{Charlier {et~al.}(2021)Charlier, Feydy, Glaunès, Collin, \&
  Durif}]{pykeops}
Charlier, B., Feydy, J., Glaunès, J.~A., Collin, F.-D., \& Durif, G. 2021,
  Journal of Machine Learning Research, 22, 1.
\newblock \url{http://jmlr.org/papers/v22/20-275.html}

\bibitem[{{Charlot} \& {Fall}(2000)}]{Charlot2000}
{Charlot}, S., \& {Fall}, S.~M. 2000, \apj, 539, 718, \dodoi{10.1086/309250}

\bibitem[{{Choi} {et~al.}(2016){Choi}, {Dotter}, {Conroy}, {Cantiello},
  {Paxton}, \& {Johnson}}]{Choi2016}
{Choi}, J., {Dotter}, A., {Conroy}, C., {et~al.} 2016, \apj, 823, 102,
  \dodoi{10.3847/0004-637X/823/2/102}

\bibitem[{Chui \& Rangarajan(2000)}]{Chui2000}
Chui, H., \& Rangarajan, A. 2000, in Proceedings IEEE Conference on Computer
  Vision and Pattern Recognition. CVPR 2000 (Cat. No.PR00662), Vol.~2, 44--51
  vol.2, \dodoi{10.1109/CVPR.2000.854733}

\bibitem[{{Cichocki} \& {Phan}(2009)}]{Cichocki2009}
{Cichocki}, A., \& {Phan}, A.-H. 2009, IEICE Transactions on Fundamentals of
  Electronics Communications and Computer Sciences, 92, 708,
  \dodoi{10.1587/transfun.E92.A.708}

\bibitem[{{Ciuca} \& {Ting}(2022)}]{Ciuca2022}
{Ciuca}, I., \& {Ting}, Y.-S. 2022, in Machine Learning for Astrophysics, 17,
  \dodoi{10.48550/arXiv.2207.02785}

\bibitem[{{Conroy}(2013)}]{Conroy2013ARAA}
{Conroy}, C. 2013, \araa, 51, 393, \dodoi{10.1146/annurev-astro-082812-141017}

\bibitem[{{Conroy} \& {Gunn}(2010)}]{Conroy2010}
{Conroy}, C., \& {Gunn}, J.~E. 2010, \apj, 712, 833,
  \dodoi{10.1088/0004-637X/712/2/833}

\bibitem[{{Conroy} {et~al.}(2009){Conroy}, {Gunn}, \& {White}}]{Conroy2009}
{Conroy}, C., {Gunn}, J.~E., \& {White}, M. 2009, \apj, 699, 486,
  \dodoi{10.1088/0004-637X/699/1/486}

\bibitem[{{Curti} {et~al.}(2020){Curti}, {Mannucci}, {Cresci}, \&
  {Maiolino}}]{Curti2020}
{Curti}, M., {Mannucci}, F., {Cresci}, G., \& {Maiolino}, R. 2020, \mnras, 491,
  944, \dodoi{10.1093/mnras/stz2910}

\bibitem[{Cuturi(2013)}]{Cuturi2013}
Cuturi, M. 2013, in Advances in Neural Information Processing Systems, ed.
  C.~Burges, L.~Bottou, M.~Welling, Z.~Ghahramani, \& K.~Weinberger, Vol.~26
  (Curran Associates, Inc.).
\newblock
  \url{https://proceedings.neurips.cc/paper_files/paper/2013/file/af21d0c97db2e27e13572cbf59eb343d-Paper.pdf}

\bibitem[{{Dai} \& {Seljak}(2022)}]{Dai2022}
{Dai}, B., \& {Seljak}, U. 2022, \mnras, 516, 2363,
  \dodoi{10.1093/mnras/stac2010}

\bibitem[{{Dalal} {et~al.}(2023){Dalal}, {Li}, {Nicola}, {Zuntz}, {Strauss},
  {Sugiyama}, {Zhang}, {Rau}, {Mandelbaum}, {Takada}, {More}, {Miyatake},
  {Kannawadi}, {Shirasaki}, {Taniguchi}, {Takahashi}, {Osato}, {Hamana},
  {Oguri}, {Nishizawa}, {Plazas Malag{\'o}n}, {Sunayama}, {Alonso}, {Slosar},
  {Armstrong}, {Bosch}, {Komiyama}, {Lupton}, {Lust}, {MacArthur}, {Miyazaki},
  {Murayama}, {Nishimichi}, {Okura}, {Price}, {Tait}, {Tanaka}, \&
  {Wang}}]{Dalal2023}
{Dalal}, R., {Li}, X., {Nicola}, A., {et~al.} 2023, arXiv e-prints,
  arXiv:2304.00701, \dodoi{10.48550/arXiv.2304.00701}

\bibitem[{{Doi} {et~al.}(2010){Doi}, {Tanaka}, {Fukugita}, {Gunn}, {Yasuda},
  {Ivezi{\'c}}, {Brinkmann}, {de Haars}, {Kleinman}, {Krzesinski}, \& {French
  Leger}}]{Doi2010}
{Doi}, M., {Tanaka}, M., {Fukugita}, M., {et~al.} 2010, \aj, 139, 1628,
  \dodoi{10.1088/0004-6256/139/4/1628}

\bibitem[{{Dotter}(2016)}]{Dotter2016}
{Dotter}, A. 2016, \apjs, 222, 8, \dodoi{10.3847/0067-0049/222/1/8}

\bibitem[{{Driver} {et~al.}(2011){Driver}, {Hill}, {Kelvin}, {Robotham},
  {Liske}, {Norberg}, {Baldry}, {Bamford}, {Hopkins}, {Loveday}, {Peacock},
  {Andrae}, {Bland-Hawthorn}, {Brough}, {Brown}, {Cameron}, {Ching}, {Colless},
  {Conselice}, {Croom}, {Cross}, {de Propris}, {Dye}, {Drinkwater}, {Ellis},
  {Graham}, {Grootes}, {Gunawardhana}, {Jones}, {van Kampen}, {Maraston},
  {Nichol}, {Parkinson}, {Phillipps}, {Pimbblet}, {Popescu}, {Prescott},
  {Roseboom}, {Sadler}, {Sansom}, {Sharp}, {Smith}, {Taylor}, {Thomas},
  {Tuffs}, {Wijesinghe}, {Dunne}, {Frenk}, {Jarvis}, {Madore}, {Meyer},
  {Seibert}, {Staveley-Smith}, {Sutherland}, \& {Warren}}]{Driver2011}
{Driver}, S.~P., {Hill}, D.~T., {Kelvin}, L.~S., {et~al.} 2011, \mnras, 413,
  971, \dodoi{10.1111/j.1365-2966.2010.18188.x}

\bibitem[{{Driver} {et~al.}(2016){Driver}, {Wright}, {Andrews}, {Davies},
  {Kafle}, {Lange}, {Moffett}, {Mannering}, {Robotham}, {Vinsen}, {Alpaslan},
  {Andrae}, {Baldry}, {Bauer}, {Bamford}, {Bland-Hawthorn}, {Bourne}, {Brough},
  {Brown}, {Cluver}, {Croom}, {Colless}, {Conselice}, {da Cunha}, {De Propris},
  {Drinkwater}, {Dunne}, {Eales}, {Edge}, {Frenk}, {Graham}, {Grootes},
  {Holwerda}, {Hopkins}, {Ibar}, {van Kampen}, {Kelvin}, {Jarrett}, {Jones},
  {Lara-Lopez}, {Liske}, {Lopez-Sanchez}, {Loveday}, {Maddox}, {Madore},
  {Mahajan}, {Meyer}, {Norberg}, {Penny}, {Phillipps}, {Popescu}, {Tuffs},
  {Peacock}, {Pimbblet}, {Prescott}, {Rowlands}, {Sansom}, {Seibert}, {Smith},
  {Sutherland}, {Taylor}, {Valiante}, {Vazquez-Mata}, {Wang}, {Wilkins}, \&
  {Williams}}]{Driver2016}
{Driver}, S.~P., {Wright}, A.~H., {Andrews}, S.~K., {et~al.} 2016, \mnras, 455,
  3911, \dodoi{10.1093/mnras/stv2505}

\bibitem[{{Durkan} {et~al.}(2019){Durkan}, {Bekasov}, {Murray}, \&
  {Papamakarios}}]{Durkan2019}
{Durkan}, C., {Bekasov}, A., {Murray}, I., \& {Papamakarios}, G. 2019, arXiv
  e-prints, arXiv:1906.04032, \dodoi{10.48550/arXiv.1906.04032}

\bibitem[{{Edge} {et~al.}(2013){Edge}, {Sutherland}, {Kuijken}, {Driver},
  {McMahon}, {Eales}, \& {Emerson}}]{VIKING2013}
{Edge}, A., {Sutherland}, W., {Kuijken}, K., {et~al.} 2013, The Messenger, 154,
  32

\bibitem[{{Elfwing} {et~al.}(2017){Elfwing}, {Uchibe}, \& {Doya}}]{Elfwing2017}
{Elfwing}, S., {Uchibe}, E., \& {Doya}, K. 2017, arXiv e-prints,
  arXiv:1702.03118.
\newblock \doarXiv{1702.03118}

\bibitem[{Feydy {et~al.}(2019)Feydy, S{\'e}journ{\'e}, Vialard, Amari, Trouve,
  \& Peyr{\'e}}]{geomloss}
Feydy, J., S{\'e}journ{\'e}, T., Vialard, F.-X., {et~al.} 2019, in The 22nd
  International Conference on Artificial Intelligence and Statistics,
  2681--2690

\bibitem[{{Genel} {et~al.}(2014){Genel}, {Vogelsberger}, {Springel}, {Sijacki},
  {Nelson}, {Snyder}, {Rodriguez-Gomez}, {Torrey}, \& {Hernquist}}]{Genel2014}
{Genel}, S., {Vogelsberger}, M., {Springel}, V., {et~al.} 2014, \mnras, 445,
  175, \dodoi{10.1093/mnras/stu1654}

\bibitem[{{Green} {et~al.}(2023){Green}, {Ting}, \& {Kamdar}}]{Green2023}
{Green}, G.~M., {Ting}, Y.-S., \& {Kamdar}, H. 2023, \apj, 942, 26,
  \dodoi{10.3847/1538-4357/aca3a7}

\bibitem[{{Greenberg} {et~al.}(2019){Greenberg}, {Nonnenmacher}, \&
  {Macke}}]{Greenberg2019}
{Greenberg}, D.~S., {Nonnenmacher}, M., \& {Macke}, J.~H. 2019, arXiv e-prints,
  arXiv:1905.07488, \dodoi{10.48550/arXiv.1905.07488}

\bibitem[{{Greene} {et~al.}(2022){Greene}, {Bezanson}, {Ouchi}, {Silverman}, \&
  {the PFS Galaxy Evolution Working Group}}]{GreenePFS2022}
{Greene}, J., {Bezanson}, R., {Ouchi}, M., {Silverman}, J., \& {the PFS Galaxy
  Evolution Working Group}. 2022, arXiv e-prints, arXiv:2206.14908,
  \dodoi{10.48550/arXiv.2206.14908}

\bibitem[{{Gunn} {et~al.}(2006){Gunn}, {Siegmund}, {Mannery}, {Owen}, {Hull},
  {Leger}, {Carey}, {Knapp}, {York}, {Boroski}, {Kent}, {Lupton}, {Rockosi},
  {Evans}, {Waddell}, {Anderson}, {Annis}, {Barentine}, {Bartoszek}, {Bastian},
  {Bracker}, {Brewington}, {Briegel}, {Brinkmann}, {Brown}, {Carr},
  {Czarapata}, {Drennan}, {Dombeck}, {Federwitz}, {Gillespie}, {Gonzales},
  {Hansen}, {Harvanek}, {Hayes}, {Jordan}, {Kinney}, {Klaene}, {Kleinman},
  {Kron}, {Kresinski}, {Lee}, {Limmongkol}, {Lindenmeyer}, {Long}, {Loomis},
  {McGehee}, {Mantsch}, {Neilsen}, {Neswold}, {Newman}, {Nitta}, {Peoples},
  {Pier}, {Prieto}, {Prosapio}, {Rivetta}, {Schneider}, {Snedden}, \&
  {Wang}}]{Gunn2006}
{Gunn}, J.~E., {Siegmund}, W.~A., {Mannery}, E.~J., {et~al.} 2006, \aj, 131,
  2332, \dodoi{10.1086/500975}

\bibitem[{{Hahn} \& {Melchior}(2022)}]{Hahn2022sedflow}
{Hahn}, C., \& {Melchior}, P. 2022, \apj, 938, 11,
  \dodoi{10.3847/1538-4357/ac7b84}

\bibitem[{{Hahn} {et~al.}(2023{\natexlab{a}}){Hahn}, {Aguilar}, {Alam},
  {Ahlen}, {Brooks}, {Cole}, {de la Macorra}, {Doel}, {Font-Ribera},
  {Forero-Romero}, {Gontcho}, {Honscheid}, {Huang}, {Kisner}, {Kremin},
  {Landriau}, {Manera}, {Meisner}, {Miquel}, {Moustakas}, {Nie}, {Poppett},
  {Rossi}, {Saintonge}, {Sanchez}, {Saulder}, {Schubnell}, {Seo}, {Siudek},
  {Speranza}, {Tarl{\'e}}, {Weaver}, {Wechsler}, {Yuan}, {Zhou}, \&
  {Zou}}]{Hahn2023SMF}
{Hahn}, C., {Aguilar}, J.~N., {Alam}, S., {et~al.} 2023{\natexlab{a}}, arXiv
  e-prints, arXiv:2306.06318, \dodoi{10.48550/arXiv.2306.06318}

\bibitem[{{Hahn} {et~al.}(2023{\natexlab{b}}){Hahn}, {Kwon}, {Tojeiro},
  {Siudek}, {Canning}, {Mezcua}, {Tinker}, {Brooks}, {Doel}, {Fanning},
  {Gazta{\~n}aga}, {Kehoe}, {Landriau}, {Meisner}, {Moustakas}, {Poppett},
  {Tarle}, {Weiner}, \& {Zou}}]{Hahn2022PROVABGS}
{Hahn}, C., {Kwon}, K.~J., {Tojeiro}, R., {et~al.} 2023{\natexlab{b}}, \apj,
  945, 16, \dodoi{10.3847/1538-4357/ac8983}

\bibitem[{{Hahn} {et~al.}(2023{\natexlab{c}}){Hahn}, {Wilson}, {Ruiz-Macias},
  {Cole}, {Weinberg}, {Moustakas}, {Kremin}, {Tinker}, {Smith}, {Wechsler},
  {Ahlen}, {Alam}, {Bailey}, {Brooks}, {Cooper}, {Davis}, {Dawson}, {Dey},
  {Dey}, {Eftekharzadeh}, {Eisenstein}, {Fanning}, {Forero-Romero}, {Frenk},
  {Gazta{\~n}aga}, {Gontcho A Gontcho}, {Guy}, {Honscheid}, {Ishak}, {Juneau},
  {Kehoe}, {Kisner}, {Lan}, {Landriau}, {Le Guillou}, {Levi}, {Magneville},
  {Martini}, {Meisner}, {Myers}, {Nie}, {Norberg}, {Palanque-Delabrouille},
  {Percival}, {Poppett}, {Prada}, {Raichoor}, {Ross}, {Safonova}, {Saulder},
  {Schlafly}, {Schlegel}, {Sierra-Porta}, {Tarle}, {Weaver}, {Y{\`e}che},
  {Zarrouk}, {Zhou}, {Zhou}, \& {Zou}}]{DESI-BGS}
{Hahn}, C., {Wilson}, M.~J., {Ruiz-Macias}, O., {et~al.} 2023{\natexlab{c}},
  \aj, 165, 253, \dodoi{10.3847/1538-3881/accff8}

\bibitem[{Harris {et~al.}(2020)Harris, Millman, van~der Walt, Gommers,
  Virtanen, Cournapeau, Wieser, Taylor, Berg, Smith, Kern, Picus, Hoyer, van
  Kerkwijk, Brett, Haldane, del R{'{\i}}o, Wiebe, Peterson,
  G{'{e}}rard-Marchant, Sheppard, Reddy, Weckesser, Abbasi, Gohlke, \&
  Oliphant}]{Numpy}
Harris, C.~R., Millman, K.~J., van~der Walt, S.~J., {et~al.} 2020, Nature, 585,
  357, \dodoi{10.1038/s41586-020-2649-2}

\bibitem[{{Hearin} {et~al.}(2023){Hearin}, {Chaves-Montero}, {Alarcon},
  {Becker}, \& {Benson}}]{Hearin2021}
{Hearin}, A.~P., {Chaves-Montero}, J., {Alarcon}, A., {Becker}, M.~R., \&
  {Benson}, A. 2023, \mnras, 521, 1741, \dodoi{10.1093/mnras/stad456}

\bibitem[{{Holzschuh} {et~al.}(2022){Holzschuh}, {O'Riordan}, {Vegetti},
  {Rodriguez-Gomez}, \& {Thuerey}}]{Holzschuh2022}
{Holzschuh}, B.~J., {O'Riordan}, C.~M., {Vegetti}, S., {Rodriguez-Gomez}, V.,
  \& {Thuerey}, N. 2022, \mnras, 515, 652, \dodoi{10.1093/mnras/stac1188}

\bibitem[{{Hunter}(2007)}]{matplotlib}
{Hunter}, J.~D. 2007, Computing in Science Engineering, 9, 90,
  \dodoi{10.1109/MCSE.2007.55}

\bibitem[{{Ivezi{\'c}} {et~al.}(2019){Ivezi{\'c}}, {Kahn}, {Tyson}, {Abel},
  {Acosta}, {Allsman}, {Alonso}, {AlSayyad}, {Anderson}, {Andrew}, {Angel},
  {Angeli}, {Ansari}, {Antilogus}, {Araujo}, {Armstrong}, {Arndt}, {Astier},
  {Aubourg}, {Auza}, {Axelrod}, {Bard}, {Barr}, {Barrau}, {Bartlett}, {Bauer},
  {Bauman}, {Baumont}, {Bechtol}, {Bechtol}, {Becker}, {Becla}, {Beldica},
  {Bellavia}, {Bianco}, {Biswas}, {Blanc}, {Blazek}, {Blandford}, {Bloom},
  {Bogart}, {Bond}, {Booth}, {Borgland}, {Borne}, {Bosch}, {Boutigny},
  {Brackett}, {Bradshaw}, {Brandt}, {Brown}, {Bullock}, {Burchat}, {Burke},
  {Cagnoli}, {Calabrese}, {Callahan}, {Callen}, {Carlin}, {Carlson},
  {Chandrasekharan}, {Charles-Emerson}, {Chesley}, {Cheu}, {Chiang}, {Chiang},
  {Chirino}, {Chow}, {Ciardi}, {Claver}, {Cohen-Tanugi}, {Cockrum}, {Coles},
  {Connolly}, {Cook}, {Cooray}, {Covey}, {Cribbs}, {Cui}, {Cutri}, {Daly},
  {Daniel}, {Daruich}, {Daubard}, {Daues}, {Dawson}, {Delgado}, {Dellapenna},
  {de Peyster}, {de Val-Borro}, {Digel}, {Doherty}, {Dubois},
  {Dubois-Felsmann}, {Durech}, {Economou}, {Eifler}, {Eracleous}, {Emmons},
  {Fausti Neto}, {Ferguson}, {Figueroa}, {Fisher-Levine}, {Focke}, {Foss},
  {Frank}, {Freemon}, {Gangler}, {Gawiser}, {Geary}, {Gee}, {Geha}, {Gessner},
  {Gibson}, {Gilmore}, {Glanzman}, {Glick}, {Goldina}, {Goldstein}, {Goodenow},
  {Graham}, {Gressler}, {Gris}, {Guy}, {Guyonnet}, {Haller}, {Harris},
  {Hascall}, {Haupt}, {Hernandez}, {Herrmann}, {Hileman}, {Hoblitt}, {Hodgson},
  {Hogan}, {Howard}, {Huang}, {Huffer}, {Ingraham}, {Innes}, {Jacoby}, {Jain},
  {Jammes}, {Jee}, {Jenness}, {Jernigan}, {Jevremovi{\'c}}, {Johns}, {Johnson},
  {Johnson}, {Jones}, {Juramy-Gilles}, {Juri{\'c}}, {Kalirai}, {Kallivayalil},
  {Kalmbach}, {Kantor}, {Karst}, {Kasliwal}, {Kelly}, {Kessler}, {Kinnison},
  {Kirkby}, {Knox}, {Kotov}, {Krabbendam}, {Krughoff}, {Kub{\'a}nek},
  {Kuczewski}, {Kulkarni}, {Ku}, {Kurita}, {Lage}, {Lambert}, {Lange},
  {Langton}, {Le Guillou}, {Levine}, {Liang}, {Lim}, {Lintott}, {Long},
  {Lopez}, {Lotz}, {Lupton}, {Lust}, {MacArthur}, {Mahabal}, {Mandelbaum},
  {Markiewicz}, {Marsh}, {Marshall}, {Marshall}, {May}, {McKercher}, {McQueen},
  {Meyers}, {Migliore}, {Miller}, {Mills}, {Miraval}, {Moeyens}, {Moolekamp},
  {Monet}, {Moniez}, {Monkewitz}, {Montgomery}, {Morrison}, {Mueller},
  {Muller}, {Mu{\~n}oz Arancibia}, {Neill}, {Newbry}, {Nief}, {Nomerotski},
  {Nordby}, {O'Connor}, {Oliver}, {Olivier}, {Olsen}, {O'Mullane}, {Ortiz},
  {Osier}, {Owen}, {Pain}, {Palecek}, {Parejko}, {Parsons}, {Pease},
  {Peterson}, {Peterson}, {Petravick}, {Libby Petrick}, {Petry},
  {Pierfederici}, {Pietrowicz}, {Pike}, {Pinto}, {Plante}, {Plate}, {Plutchak},
  {Price}, {Prouza}, {Radeka}, {Rajagopal}, {Rasmussen}, {Regnault}, {Reil},
  {Reiss}, {Reuter}, {Ridgway}, {Riot}, {Ritz}, {Robinson}, {Roby}, {Roodman},
  {Rosing}, {Roucelle}, {Rumore}, {Russo}, {Saha}, {Sassolas}, {Schalk},
  {Schellart}, {Schindler}, {Schmidt}, {Schneider}, {Schneider}, {Schoening},
  {Schumacher}, {Schwamb}, {Sebag}, {Selvy}, {Sembroski}, {Seppala}, {Serio},
  {Serrano}, {Shaw}, {Shipsey}, {Sick}, {Silvestri}, {Slater}, {Smith},
  {Smith}, {Sobhani}, {Soldahl}, {Storrie-Lombardi}, {Stover}, {Strauss},
  {Street}, {Stubbs}, {Sullivan}, {Sweeney}, {Swinbank}, {Szalay}, {Takacs},
  {Tether}, {Thaler}, {Thayer}, {Thomas}, {Thornton}, {Thukral}, {Tice},
  {Trilling}, {Turri}, {Van Berg}, {Vanden Berk}, {Vetter}, {Virieux},
  {Vucina}, {Wahl}, {Walkowicz}, {Walsh}, {Walter}, {Wang}, {Wang}, {Warner},
  {Wiecha}, {Willman}, {Winters}, {Wittman}, {Wolff}, {Wood-Vasey}, {Wu},
  {Xin}, {Yoachim}, \& {Zhan}}]{LSST2019}
{Ivezi{\'c}}, {\v{Z}}., {Kahn}, S.~M., {Tyson}, J.~A., {et~al.} 2019, \apj,
  873, 111, \dodoi{10.3847/1538-4357/ab042c}

\bibitem[{Johnson {et~al.}(2021)Johnson, Foreman-Mackey, Sick, Leja, Byler,
  Walmsley, Tollerud, Leung, \& Scott}]{pyfsps}
Johnson, B., Foreman-Mackey, D., Sick, J., {et~al.} 2021, dfm/python-fsps:
  python-fsps v0.4.1rc1, v0.4.1rc1,  Zenodo, \dodoi{10.5281/zenodo.4737461}

\bibitem[{Johnson(2021)}]{sedpy}
Johnson, B.~D. 2021, bd-j/sedpy: sedpy v0.2.0, v0.2.0,  Zenodo,
  \dodoi{10.5281/zenodo.4582723}

\bibitem[{{Johnson} {et~al.}(2021){Johnson}, {Leja}, {Conroy}, \&
  {Speagle}}]{Johnson2021}
{Johnson}, B.~D., {Leja}, J., {Conroy}, C., \& {Speagle}, J.~S. 2021, \apjs,
  254, 22, \dodoi{10.3847/1538-4365/abef67}

\bibitem[{Jones {et~al.}(2001)Jones, Oliphant, Peterson, {et~al.}}]{scipy}
Jones, E., Oliphant, T., Peterson, P., {et~al.} 2001, {SciPy}: Open source
  scientific tools for {Python}.
\newblock \url{http://www.scipy.org/}

\bibitem[{{Khullar} {et~al.}(2022){Khullar}, {Nord}, {{\'C}iprijanovi{\'c}},
  {Poh}, \& {Xu}}]{Khullar2022}
{Khullar}, G., {Nord}, B., {{\'C}iprijanovi{\'c}}, A., {Poh}, J., \& {Xu}, F.
  2022, Machine Learning: Science and Technology, 3, 04LT04,
  \dodoi{10.1088/2632-2153/ac98f4}

\bibitem[{{Kingma} \& {Ba}(2014)}]{Adam}
{Kingma}, D.~P., \& {Ba}, J. 2014, arXiv e-prints, arXiv:1412.6980,
  \dodoi{10.48550/arXiv.1412.6980}

\bibitem[{{Kobyzev} {et~al.}(2019){Kobyzev}, {Prince}, \&
  {Brubaker}}]{Kobyzev2019}
{Kobyzev}, I., {Prince}, S. J.~D., \& {Brubaker}, M.~A. 2019, arXiv e-prints,
  arXiv:1908.09257, \dodoi{10.48550/arXiv.1908.09257}

\bibitem[{{Kolouri} {et~al.}(2018){Kolouri}, {Pope}, {Martin}, \&
  {Rohde}}]{Kolouri2018}
{Kolouri}, S., {Pope}, P.~E., {Martin}, C.~E., \& {Rohde}, G.~K. 2018, arXiv
  e-prints, arXiv:1804.01947, \dodoi{10.48550/arXiv.1804.01947}

\bibitem[{{Labb{\'e}} {et~al.}(2023){Labb{\'e}}, {van Dokkum}, {Nelson},
  {Bezanson}, {Suess}, {Leja}, {Brammer}, {Whitaker}, {Mathews}, {Stefanon}, \&
  {Wang}}]{Labbe2023}
{Labb{\'e}}, I., {van Dokkum}, P., {Nelson}, E., {et~al.} 2023, \nat, 616, 266,
  \dodoi{10.1038/s41586-023-05786-2}

\bibitem[{{Lee} \& {Seung}(1999)}]{Lee1999}
{Lee}, D.~D., \& {Seung}, H.~S. 1999, \nat, 401, 788, \dodoi{10.1038/44565}

\bibitem[{{Leistedt} {et~al.}(2016){Leistedt}, {Mortlock}, \&
  {Peiris}}]{Leistedt2016}
{Leistedt}, B., {Mortlock}, D.~J., \& {Peiris}, H.~V. 2016, \mnras, 460, 4258,
  \dodoi{10.1093/mnras/stw1304}

\bibitem[{{Leja} {et~al.}(2019{\natexlab{a}}){Leja}, {Carnall}, {Johnson},
  {Conroy}, \& {Speagle}}]{Leja2019}
{Leja}, J., {Carnall}, A.~C., {Johnson}, B.~D., {Conroy}, C., \& {Speagle},
  J.~S. 2019{\natexlab{a}}, \apj, 876, 3, \dodoi{10.3847/1538-4357/ab133c}

\bibitem[{{Leja} {et~al.}(2019{\natexlab{b}}){Leja}, {Johnson}, {Conroy}, {van
  Dokkum}, {Speagle}, {Brammer}, {Momcheva}, {Skelton}, {Whitaker}, {Franx}, \&
  {Nelson}}]{Leja2019b}
{Leja}, J., {Johnson}, B.~D., {Conroy}, C., {et~al.} 2019{\natexlab{b}}, \apj,
  877, 140, \dodoi{10.3847/1538-4357/ab1d5a}

\bibitem[{{Lejeune} {et~al.}(1997){Lejeune}, {Cuisinier}, \&
  {Buser}}]{Lejeune1997}
{Lejeune}, T., {Cuisinier}, F., \& {Buser}, R. 1997, \aaps, 125, 229,
  \dodoi{10.1051/aas:1997373}

\bibitem[{{Lejeune} {et~al.}(1998){Lejeune}, {Cuisinier}, \&
  {Buser}}]{Lejeune1998}
---. 1998, \aaps, 130, 65, \dodoi{10.1051/aas:1998405}

\bibitem[{Li {et~al.}(2023)Li, Melchior, Hahn, \& Huang}]{li_2023_10094993}
Li, J., Melchior, P., Hahn, C., \& Huang, S. 2023, {PopSED: Population-Level
  Inference for Galaxy Properties}, v0.0.6,  Zenodo,
  \dodoi{10.5281/zenodo.10094993}

\bibitem[{{Liang} {et~al.}(2023){Liang}, {Melchior}, {Lu}, {Goulding}, \&
  {Ward}}]{Liang2023}
{Liang}, Y., {Melchior}, P., {Lu}, S., {Goulding}, A., \& {Ward}, C. 2023, \aj,
  166, 75, \dodoi{10.3847/1538-3881/ace100}

\bibitem[{{Luo} {et~al.}(2023){Luo}, {Leauthaud}, {Greene}, {Huang},
  {Kado-Fong}, {Danieli}, {Li}, {Li}, {Blanco}, {Wasleske}, {Wick}, {Mintz},
  {Guan}, {Peter}, {Baldassare}, {Brooks}, {Banerjee}, {Bhattacharyya}, {Cai},
  {Chen}, {Gunn}, {Johnson}, {Kelvin}, {Li}, {Lin}, {Lupton}, {Mace}, {Medina},
  {Read}, {Cordova Rosado}, \& {Seifert}}]{Luo2023}
{Luo}, Y., {Leauthaud}, A., {Greene}, J., {et~al.} 2023, arXiv e-prints,
  arXiv:2305.19310, \dodoi{10.48550/arXiv.2305.19310}

\bibitem[{{Lupton} {et~al.}(1999){Lupton}, {Gunn}, \& {Szalay}}]{Lupton1999}
{Lupton}, R.~H., {Gunn}, J.~E., \& {Szalay}, A.~S. 1999, \aj, 118, 1406,
  \dodoi{10.1086/301004}

\bibitem[{{Malz} \& {Hogg}(2020)}]{Malz2020}
{Malz}, A.~I., \& {Hogg}, D.~W. 2020, arXiv e-prints, arXiv:2007.12178,
  \dodoi{10.48550/arXiv.2007.12178}

\bibitem[{{Mandelbaum}(2018)}]{Mandelbaum2018}
{Mandelbaum}, R. 2018, \araa, 56, 393,
  \dodoi{10.1146/annurev-astro-081817-051928}

\bibitem[{{McGaugh} {et~al.}(2017){McGaugh}, {Schombert}, \&
  {Lelli}}]{McGaugh2017}
{McGaugh}, S.~S., {Schombert}, J.~M., \& {Lelli}, F. 2017, \apj, 851, 22,
  \dodoi{10.3847/1538-4357/aa9790}

\bibitem[{{Nelson} {et~al.}(2015){Nelson}, {Pillepich}, {Genel},
  {Vogelsberger}, {Springel}, {Torrey}, {Rodriguez-Gomez}, {Sijacki}, {Snyder},
  {Griffen}, {Marinacci}, {Blecha}, {Sales}, {Xu}, \& {Hernquist}}]{Nelson2015}
{Nelson}, D., {Pillepich}, A., {Genel}, S., {et~al.} 2015, Astronomy and
  Computing, 13, 12, \dodoi{10.1016/j.ascom.2015.09.003}

\bibitem[{{Newman} \& {Gruen}(2022)}]{Newman2022}
{Newman}, J.~A., \& {Gruen}, D. 2022, \araa, 60, 363,
  \dodoi{10.1146/annurev-astro-032122-014611}

\bibitem[{{Noll} {et~al.}(2009){Noll}, {Burgarella}, {Giovannoli}, {Buat},
  {Marcillac}, \& {Mu{\~n}oz-Mateos}}]{Noll2009}
{Noll}, S., {Burgarella}, D., {Giovannoli}, E., {et~al.} 2009, \aap, 507, 1793,
  \dodoi{10.1051/0004-6361/200912497}

\bibitem[{{Oke} \& {Gunn}(1983)}]{Oke1983}
{Oke}, J.~B., \& {Gunn}, J.~E. 1983, \apj, 266, 713, \dodoi{10.1086/160817}

\bibitem[{{Papamakarios} {et~al.}(2017){Papamakarios}, {Pavlakou}, \&
  {Murray}}]{Papamakarios2017}
{Papamakarios}, G., {Pavlakou}, T., \& {Murray}, I. 2017, arXiv e-prints,
  arXiv:1705.07057, \dodoi{10.48550/arXiv.1705.07057}

\bibitem[{Paszke {et~al.}(2019)Paszke, Gross, Massa, Lerer, Bradbury, Chanan,
  Killeen, Lin, Gimelshein, Antiga, Desmaison, Kopf, Yang, DeVito, Raison,
  Tejani, Chilamkurthy, Steiner, Fang, Bai, \& Chintala}]{pytorch}
Paszke, A., Gross, S., Massa, F., {et~al.} 2019, in Advances in Neural
  Information Processing Systems 32, ed. H.~Wallach, H.~Larochelle,
  A.~Beygelzimer, F.~d\textquotesingle Alch\'{e}-Buc, E.~Fox, \& R.~Garnett
  (Curran Associates, Inc.), 8024--8035.
\newblock
  \url{http://papers.neurips.cc/paper/9015-pytorch-an-imperative-style-high-performance-deep-learning-library.pdf}

\bibitem[{{Peyr{\'e}} \& {Cuturi}(2018)}]{Peyre2018}
{Peyr{\'e}}, G., \& {Cuturi}, M. 2018, arXiv e-prints, arXiv:1803.00567,
  \dodoi{10.48550/arXiv.1803.00567}

\bibitem[{{Planck Collaboration} {et~al.}(2016){Planck Collaboration}, {Ade},
  {Aghanim}, {Arnaud}, {Ashdown}, {Aumont}, {Baccigalupi}, {Banday},
  {Barreiro}, {Bartlett}, {Bartolo}, {Battaner}, {Battye}, {Benabed},
  {Beno{\^\i}t}, {Benoit-L{\'e}vy}, {Bernard}, {Bersanelli}, {Bielewicz},
  {Bock}, {Bonaldi}, {Bonavera}, {Bond}, {Borrill}, {Bouchet}, {Boulanger},
  {Bucher}, {Burigana}, {Butler}, {Calabrese}, {Cardoso}, {Catalano},
  {Challinor}, {Chamballu}, {Chary}, {Chiang}, {Chluba}, {Christensen},
  {Church}, {Clements}, {Colombi}, {Colombo}, {Combet}, {Coulais}, {Crill},
  {Curto}, {Cuttaia}, {Danese}, {Davies}, {Davis}, {de Bernardis}, {de Rosa},
  {de Zotti}, {Delabrouille}, {D{\'e}sert}, {Di Valentino}, {Dickinson},
  {Diego}, {Dolag}, {Dole}, {Donzelli}, {Dor{\'e}}, {Douspis}, {Ducout},
  {Dunkley}, {Dupac}, {Efstathiou}, {Elsner}, {En{\ss}lin}, {Eriksen},
  {Farhang}, {Fergusson}, {Finelli}, {Forni}, {Frailis}, {Fraisse},
  {Franceschi}, {Frejsel}, {Galeotta}, {Galli}, {Ganga}, {Gauthier}, {Gerbino},
  {Ghosh}, {Giard}, {Giraud-H{\'e}raud}, {Giusarma}, {Gjerl{\o}w},
  {Gonz{\'a}lez-Nuevo}, {G{\'o}rski}, {Gratton}, {Gregorio}, {Gruppuso},
  {Gudmundsson}, {Hamann}, {Hansen}, {Hanson}, {Harrison}, {Helou},
  {Henrot-Versill{\'e}}, {Hern{\'a}ndez-Monteagudo}, {Herranz}, {Hildebrandt},
  {Hivon}, {Hobson}, {Holmes}, {Hornstrup}, {Hovest}, {Huang}, {Huffenberger},
  {Hurier}, {Jaffe}, {Jaffe}, {Jones}, {Juvela}, {Keih{\"a}nen}, {Keskitalo},
  {Kisner}, {Kneissl}, {Knoche}, {Knox}, {Kunz}, {Kurki-Suonio}, {Lagache},
  {L{\"a}hteenm{\"a}ki}, {Lamarre}, {Lasenby}, {Lattanzi}, {Lawrence}, {Leahy},
  {Leonardi}, {Lesgourgues}, {Levrier}, {Lewis}, {Liguori}, {Lilje},
  {Linden-V{\o}rnle}, {L{\'o}pez-Caniego}, {Lubin}, {Mac{\'\i}as-P{\'e}rez},
  {Maggio}, {Maino}, {Mandolesi}, {Mangilli}, {Marchini}, {Maris}, {Martin},
  {Martinelli}, {Mart{\'\i}nez-Gonz{\'a}lez}, {Masi}, {Matarrese}, {McGehee},
  {Meinhold}, {Melchiorri}, {Melin}, {Mendes}, {Mennella}, {Migliaccio},
  {Millea}, {Mitra}, {Miville-Desch{\^e}nes}, {Moneti}, {Montier}, {Morgante},
  {Mortlock}, {Moss}, {Munshi}, {Murphy}, {Naselsky}, {Nati}, {Natoli},
  {Netterfield}, {N{\o}rgaard-Nielsen}, {Noviello}, {Novikov}, {Novikov},
  {Oxborrow}, {Paci}, {Pagano}, {Pajot}, {Paladini}, {Paoletti}, {Partridge},
  {Pasian}, {Patanchon}, {Pearson}, {Perdereau}, {Perotto}, {Perrotta},
  {Pettorino}, {Piacentini}, {Piat}, {Pierpaoli}, {Pietrobon}, {Plaszczynski},
  {Pointecouteau}, {Polenta}, {Popa}, {Pratt}, {Pr{\'e}zeau}, {Prunet},
  {Puget}, {Rachen}, {Reach}, {Rebolo}, {Reinecke}, {Remazeilles}, {Renault},
  {Renzi}, {Ristorcelli}, {Rocha}, {Rosset}, {Rossetti}, {Roudier},
  {Rouill{\'e} d'Orfeuil}, {Rowan-Robinson}, {Rubi{\~n}o-Mart{\'\i}n},
  {Rusholme}, {Said}, {Salvatelli}, {Salvati}, {Sandri}, {Santos},
  {Savelainen}, {Savini}, {Scott}, {Seiffert}, {Serra}, {Shellard}, {Spencer},
  {Spinelli}, {Stolyarov}, {Stompor}, {Sudiwala}, {Sunyaev}, {Sutton},
  {Suur-Uski}, {Sygnet}, {Tauber}, {Terenzi}, {Toffolatti}, {Tomasi},
  {Tristram}, {Trombetti}, {Tucci}, {Tuovinen}, {T{\"u}rler}, {Umana},
  {Valenziano}, {Valiviita}, {Van Tent}, {Vielva}, {Villa}, {Wade}, {Wandelt},
  {Wehus}, {White}, {White}, {Wilkinson}, {Yvon}, {Zacchei}, \&
  {Zonca}}]{Planck15}
{Planck Collaboration}, {Ade}, P.~A.~R., {Aghanim}, N., {et~al.} 2016, \aap,
  594, A13, \dodoi{10.1051/0004-6361/201525830}

\bibitem[{{Racca} {et~al.}(2016){Racca}, {Laureijs}, {Stagnaro}, {Salvignol},
  {Lorenzo Alvarez}, {Saavedra Criado}, {Gaspar Venancio}, {Short}, {Strada},
  {B{\"o}nke}, {Colombo}, {Calvi}, {Maiorano}, {Piersanti}, {Prezelus},
  {Rosato}, {Pinel}, {Rozemeijer}, {Lesna}, {Musi}, {Sias}, {Anselmi},
  {Cazaubiel}, {Vaillon}, {Mellier}, {Amiaux}, {Berth{\'e}}, {Sauvage},
  {Azzollini}, {Cropper}, {Pottinger}, {Jahnke}, {Ealet}, {Maciaszek},
  {Pasian}, {Zacchei}, {Scaramella}, {Hoar}, {Kohley}, {Vavrek}, {Rudolph}, \&
  {Schmidt}}]{Euclid2016}
{Racca}, G.~D., {Laureijs}, R., {Stagnaro}, L., {et~al.} 2016, in Society of
  Photo-Optical Instrumentation Engineers (SPIE) Conference Series, Vol. 9904,
  Space Telescopes and Instrumentation 2016: Optical, Infrared, and Millimeter
  Wave, ed. H.~A. {MacEwen}, G.~G. {Fazio}, M.~{Lystrup}, N.~{Batalha},
  N.~{Siegler}, \& E.~C. {Tong}, 99040O, \dodoi{10.1117/12.2230762}

\bibitem[{{Ramachandran} {et~al.}(2017){Ramachandran}, {Zoph}, \&
  {Le}}]{Ramachandran2017}
{Ramachandran}, P., {Zoph}, B., \& {Le}, Q.~V. 2017, arXiv e-prints,
  arXiv:1710.05941.
\newblock \doarXiv{1710.05941}

\bibitem[{Ramdas {et~al.}(2017)Ramdas, Trillos, \& Cuturi}]{Ramdas2017}
Ramdas, A., Trillos, N.~G., \& Cuturi, M. 2017, Entropy, 19,
  \dodoi{10.3390/e19020047}

\bibitem[{{Renzini} \& {Peng}(2015)}]{Renzini2015}
{Renzini}, A., \& {Peng}, Y.-j. 2015, \apjl, 801, L29,
  \dodoi{10.1088/2041-8205/801/2/L29}

\bibitem[{{S{\'a}nchez} {et~al.}(2019){S{\'a}nchez}, {Avila-Reese},
  {Rodr{\'\i}guez-Puebla}, {Ibarra-Medel}, {Calette}, {Bershady},
  {Hern{\'a}ndez-Toledo}, {Pan}, \& {Bizyaev}}]{Sanchze2019}
{S{\'a}nchez}, S.~F., {Avila-Reese}, V., {Rodr{\'\i}guez-Puebla}, A., {et~al.}
  2019, \mnras, 482, 1557, \dodoi{10.1093/mnras/sty2730}

\bibitem[{{S{\'a}nchez-Bl{\'a}zquez} {et~al.}(2006){S{\'a}nchez-Bl{\'a}zquez},
  {Peletier}, {Jim{\'e}nez-Vicente}, {Cardiel}, {Cenarro},
  {Falc{\'o}n-Barroso}, {Gorgas}, {Selam}, \& {Vazdekis}}]{MILES}
{S{\'a}nchez-Bl{\'a}zquez}, P., {Peletier}, R.~F., {Jim{\'e}nez-Vicente}, J.,
  {et~al.} 2006, \mnras, 371, 703, \dodoi{10.1111/j.1365-2966.2006.10699.x}

\bibitem[{Sinkhorn(1964)}]{Sinkhorn}
Sinkhorn, R. 1964, The Annals of Mathematical Statistics, 35, 876 ,
  \dodoi{10.1214/aoms/1177703591}

\bibitem[{{Smith} \& {Topin}(2017)}]{Smith2017}
{Smith}, L.~N., \& {Topin}, N. 2017, arXiv e-prints, arXiv:1708.07120,
  \dodoi{10.48550/arXiv.1708.07120}

\bibitem[{{Speagle} {et~al.}(2014){Speagle}, {Steinhardt}, {Capak}, \&
  {Silverman}}]{Speagle2014}
{Speagle}, J.~S., {Steinhardt}, C.~L., {Capak}, P.~L., \& {Silverman}, J.~D.
  2014, \apjs, 214, 15, \dodoi{10.1088/0067-0049/214/2/15}

\bibitem[{{Spergel} {et~al.}(2015){Spergel}, {Gehrels}, {Baltay}, {Bennett},
  {Breckinridge}, {Donahue}, {Dressler}, {Gaudi}, {Greene}, {Guyon}, {Hirata},
  {Kalirai}, {Kasdin}, {Macintosh}, {Moos}, {Perlmutter}, {Postman},
  {Rauscher}, {Rhodes}, {Wang}, {Weinberg}, {Benford}, {Hudson}, {Jeong},
  {Mellier}, {Traub}, {Yamada}, {Capak}, {Colbert}, {Masters}, {Penny},
  {Savransky}, {Stern}, {Zimmerman}, {Barry}, {Bartusek}, {Carpenter}, {Cheng},
  {Content}, {Dekens}, {Demers}, {Grady}, {Jackson}, {Kuan}, {Kruk}, {Melton},
  {Nemati}, {Parvin}, {Poberezhskiy}, {Peddie}, {Ruffa}, {Wallace}, {Whipple},
  {Wollack}, \& {Zhao}}]{Spergel2015}
{Spergel}, D., {Gehrels}, N., {Baltay}, C., {et~al.} 2015, arXiv e-prints,
  arXiv:1503.03757, \dodoi{10.48550/arXiv.1503.03757}

\bibitem[{{Sun} {et~al.}(2023){Sun}, {Ting}, \& {Cai}}]{Sun2022}
{Sun}, Z., {Ting}, Y.-S., \& {Cai}, Z. 2023, \apjs, 269, 4,
  \dodoi{10.3847/1538-4365/acf2f1}

\bibitem[{Tabak \& Turner(2013)}]{Tabak2013}
Tabak, E., \& Turner, C. 2013, Communications on Pure and Applied Mathematics,
  66, 145, \dodoi{10.1002/cpa.21423}

\bibitem[{Tabak \& Vanden-Eijnden(2010)}]{Tabak2010}
Tabak, E., \& Vanden-Eijnden, E. 2010, Communications in Mathematical Sciences,
  8, 217, \dodoi{10.4310/CMS.2010.v8.n1.a11}

\bibitem[{{Taylor} {et~al.}(2011){Taylor}, {Hopkins}, {Baldry}, {Brown},
  {Driver}, {Kelvin}, {Hill}, {Robotham}, {Bland-Hawthorn}, {Jones}, {Sharp},
  {Thomas}, {Liske}, {Loveday}, {Norberg}, {Peacock}, {Bamford}, {Brough},
  {Colless}, {Cameron}, {Conselice}, {Croom}, {Frenk}, {Gunawardhana},
  {Kuijken}, {Nichol}, {Parkinson}, {Phillipps}, {Pimbblet}, {Popescu},
  {Prescott}, {Sutherland}, {Tuffs}, {van Kampen}, \&
  {Wijesinghe}}]{Taylor2011}
{Taylor}, E.~N., {Hopkins}, A.~M., {Baldry}, I.~K., {et~al.} 2011, \mnras, 418,
  1587, \dodoi{10.1111/j.1365-2966.2011.19536.x}

\bibitem[{Tejero-Cantero {et~al.}(2020)Tejero-Cantero, Boelts, Deistler,
  Lueckmann, Durkan, Gonçalves, Greenberg, \& Macke}]{SBI}
Tejero-Cantero, A., Boelts, J., Deistler, M., {et~al.} 2020, Journal of Open
  Source Software, 5, 2505, \dodoi{10.21105/joss.02505}

\bibitem[{{Thorne} {et~al.}(2021){Thorne}, {Robotham}, {Davies}, {Bellstedt},
  {Driver}, {Bravo}, {Bremer}, {Holwerda}, {Hopkins}, {Lagos}, {Phillipps},
  {Siudek}, {Taylor}, \& {Wright}}]{Thorne2021}
{Thorne}, J.~E., {Robotham}, A. S.~G., {Davies}, L. J.~M., {et~al.} 2021,
  \mnras, 505, 540, \dodoi{10.1093/mnras/stab1294}

\bibitem[{{Tremonti} {et~al.}(2004){Tremonti}, {Heckman}, {Kauffmann},
  {Brinchmann}, {Charlot}, {White}, {Seibert}, {Peng}, {Schlegel}, {Uomoto},
  {Fukugita}, \& {Brinkmann}}]{Tremonti2004}
{Tremonti}, C.~A., {Heckman}, T.~M., {Kauffmann}, G., {et~al.} 2004, \apj, 613,
  898, \dodoi{10.1086/423264}

\bibitem[{{Tsizh} {et~al.}(2023){Tsizh}, {Tymchyshyn}, \& {Vazza}}]{Tsizh2023}
{Tsizh}, M., {Tymchyshyn}, V., \& {Vazza}, F. 2023, \mnras, 522, 2697,
  \dodoi{10.1093/mnras/stad1121}

\bibitem[{{Vogelsberger} {et~al.}(2014){Vogelsberger}, {Genel}, {Springel},
  {Torrey}, {Sijacki}, {Xu}, {Snyder}, {Nelson}, \&
  {Hernquist}}]{Vogelsberger2014}
{Vogelsberger}, M., {Genel}, S., {Springel}, V., {et~al.} 2014, \mnras, 444,
  1518, \dodoi{10.1093/mnras/stu1536}

\bibitem[{{Wang} {et~al.}(2023){Wang}, {Leja}, {Villar}, \&
  {Speagle}}]{Wang2023}
{Wang}, B., {Leja}, J., {Villar}, V.~A., \& {Speagle}, J.~S. 2023, arXiv
  e-prints, arXiv:2304.05281, \dodoi{10.48550/arXiv.2304.05281}

\bibitem[{{Westera} {et~al.}(2002){Westera}, {Lejeune}, {Buser}, {Cuisinier},
  \& {Bruzual}}]{Westera2002}
{Westera}, P., {Lejeune}, T., {Buser}, R., {Cuisinier}, F., \& {Bruzual}, G.
  2002, \aap, 381, 524, \dodoi{10.1051/0004-6361:20011493}

\bibitem[{{Wilson} \& {Izmailov}(2020)}]{Wilson2020}
{Wilson}, A.~G., \& {Izmailov}, P. 2020, arXiv e-prints, arXiv:2002.08791,
  \dodoi{10.48550/arXiv.2002.08791}

\bibitem[{{Wong} {et~al.}(2020){Wong}, {Contardo}, \& {Ho}}]{Wong2020}
{Wong}, K. W.~K., {Contardo}, G., \& {Ho}, S. 2020, \prd, 101, 123005,
  \dodoi{10.1103/PhysRevD.101.123005}

\bibitem[{{Wright} {et~al.}(2017){Wright}, {Robotham}, {Driver}, {Alpaslan},
  {Andrews}, {Baldry}, {Bland-Hawthorn}, {Brough}, {Brown}, {Colless}, {da
  Cunha}, {Davies}, {Graham}, {Holwerda}, {Hopkins}, {Kafle}, {Kelvin},
  {Loveday}, {Maddox}, {Meyer}, {Moffett}, {Norberg}, {Phillipps}, {Rowlands},
  {Taylor}, {Wang}, \& {Wilkins}}]{Wright2017}
{Wright}, A.~H., {Robotham}, A.~S.~G., {Driver}, S.~P., {et~al.} 2017, \mnras,
  470, 283, \dodoi{10.1093/mnras/stx1149}

\bibitem[{{Zhang} {et~al.}(2021){Zhang}, {Bloom}, {Gaudi}, {Lanusse}, {Lam}, \&
  {Lu}}]{Zhang2021}
{Zhang}, K., {Bloom}, J.~S., {Gaudi}, B.~S., {et~al.} 2021, \aj, 161, 262,
  \dodoi{10.3847/1538-3881/abf42e}

\bibitem[{{Zhang} {et~al.}(2023){Zhang}, {Green}, \& {Rix}}]{Zhang2023}
{Zhang}, X., {Green}, G.~M., \& {Rix}, H.-W. 2023, \mnras, 524, 1855,
  \dodoi{10.1093/mnras/stad1941}

\end{thebibliography}
\bibliographystyle{aasjournal}



\end{CJK*}

\end{document}